\documentclass{JINST}
\let\ifpdf\relax
\usepackage{lscape}
\usepackage{epsfig}
\usepackage{url}
\usepackage{multirow}
\usepackage{subfigure}

\title{Acrylic Target Vessels for a High-Precision Measurement of $\theta_{13}$ with the Daya Bay Antineutrino Detectors}

\author{H.~R.~Band$^a$, R.~Brown$^b$, J.~Cherwinka$^a$, J.~Cao$^c$, Y.~Chang$^d$, B.~Edwards$^e$, W.~S.~He$^f$, K.~M.~Heeger$^a$, Y.~Heng$^c$, T.~H.~Ho$^f$, Y.~B.~Hsiung$^f$, L.~Greenler$^a$, S.~H.~Kettell$^b$, C.~A.~Lewis$^a$, K.~B.~Luk$^e$, X.~Li$^c$, B.~R.~Littlejohn$^a$\thanks{Corresponding author.}, A.~Pagac$^a$, C.~H.~Wang$^f$, W.~Wang$^g$, Y.~Wang$^c$, T.~Wise$^a$, Q.~Xiao$^a$, M.~Yeh$^b$, H.~Zhuang$^c$  \\
\llap{$^a$}University of Wisconsin, Madison,\\
  1150 University Avenue, Madison, Wisconsin\\
\llap{$^b$}Brookhaven National Laboratory,\\
  Upton, NY, USA\\
\llap{$^c$}Intitute of High Energy Physics,\\
  19 Yuquan Road, Beijing, China\\
\llap{$^d$}National United University Taiwan,\\
  No 1, Lien-Da Rd, Miao-Li, Taiwan\\
\llap{$^e$}Lawrence Berkeley National Laboratory,\\
  Address, Cyclotron Rd, Berkeley, CA, USA\\
\llap{$^f$}National Taiwan University\\
  No 1, Sec 4, Roosevelt Rd, Taipei, Taiwan\\
\llap{$^g$}College of William and Mary\\
  116 Jamestown Rd. Williamsburg, Virginia\\
  E-mail: \email{littlejohn@wisc.edu}
}

\abstract{This paper describes in detail the acrylic vessels used to encapsulate the target and gamma catcher regions in the Daya Bay experiment's first pair of antineutrino detectors.  We give an overview of the design, fabrication, shipping, and installation of the acrylic vessels and their liquid overflow tanks.  The acrylic quality assurance program and vessel characterization, which measures all geometric, optical, and material properties  relevant to $\overline{\nu}_e$ detection at Daya Bay are summarized.  This paper is the technical reference for the Daya Bay acrylic vessels and can provide guidance in the design and use of acrylic components in future neutrino or dark matter experiments.}

\keywords{Daya Bay, acrylic, quality assurance, optical properties, mechanical properties, compatibility, radioactivity}

\begin{document}
\newpage
\section{Introduction}

\begin{figure}[htb]
\centering
\includegraphics[width=3 in]{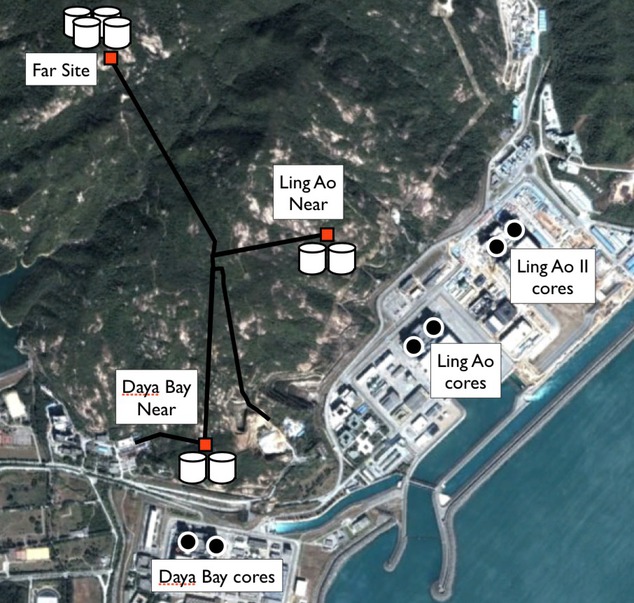}
\caption{An bird's-eye illustration of the detector and reactor locations of the Daya Bay Experiment.  Distances from the reactors to the near experiment halls are 300-600~m; distances from the reactors to the far hall are 1600-2000~m.}
\label{fig:site}
\end{figure}

The Daya Bay reactor neutrino experiment is designed to measure the neutrino mixing parameter $\sin^2 2\theta_{13}$ to better than 0.01 at 90\% confidence level via the observation of reactor-$\overline{\nu}_e$ disappearance~\cite{DayaBay:2007}.  The experiment consists of eight identical $\overline{\nu}_e$ detectors (ADs) placed at three underground experimental halls at near and far baselines from six nuclear reactor cores in a configuration as shown in Figure~\ref{fig:site}.  The Daya Bay experiment will be one of the first reactor neutrino experiments to measure relative antineutrino rates and spectra with near and far detectors at km-order baselines.

\begin{figure}[htb]
\centering
\includegraphics[trim=11cm 3cm 1cm 5cm, clip=true, width=5 in]{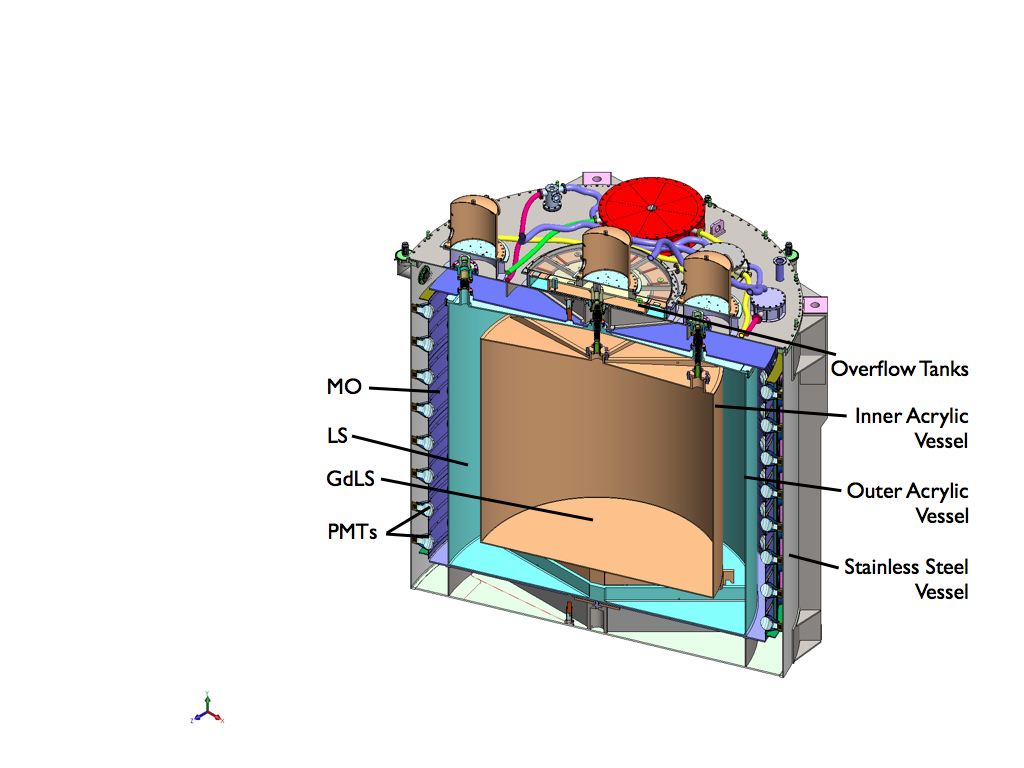}
\caption{Cutaway view of a fully constructed antineutrino detector.}
\label{fig:AD-3D}
\end{figure}

An antineutrino detector (AD), visible in Figure~\ref{fig:AD-3D}, consists of three concentric liquid zones separated by two ultraviolet-transmitting acrylic vessels.  The innermost zone, composed of Gd-doped liquid scintillator (Gd-LS), is the antineutrino target.  The middle region, filled with undoped liquid scintillator (LS), is a gamma catcher that absorbs energy from events on the edge of the target region, precluding the need for a fiducial volume cut.  The outermost liquid region, filled with mineral oil (MO), is a buffer between the photomultiplier tubes (PMTs) around the outside edge of the detector and the two central regions.  This region helps reduce radioactive backgrounds and improve detector energy resolution.  $\overline{\nu_e}$ interacting via inverse-beta decay provide a clear signal by depositing a prompt energy pulse from positron energy deposition and annihilation in either of the two inner liquid scintillator volumes and a delayed 8~MeV energy pulse from neutron capture on Gd in the target volume.  The ADs are assembled in pairs to allow for a relative comparison between detectors that are as identical as possible.  The first two ADs will be deployed and run in the Daya Bay near site experimental hall while the rest of the experiment's construction in completed.  This near site data will be used to check the performance and response of the two 'identical' ADs and examine detector and reactor systematics.

As the detector systematic uncertainty is the dominant systematic in the measurement of a near-far $\overline{\nu}_e$ flux ratio, it must be kept well below the percent level if a percent-level relative disappearance is to be measured.  Estimates of detector-related systematics for the Daya Bay experiment are shown in Table~\ref{tab:detector-sys}.  The largest of all detector systematics are those related to the number of target protons and energy cuts done to extract signal from background.  An important task in achieving these stated systematics goals is understanding and characterizing the acrylic vessels that determine many of the properties of the detector target volume.

\begin{table}[htpb]
\centering
\begin{tabular}{|l|l|c|c|c|}
\hline
\multicolumn{2}{|c|}{\multirow{2}{*}{Source of uncertainty}}
&Absolute uncertainty &\multicolumn{2}{c|}{Relative Uncertainty (\%)} \\
\cline{4-5} 
\multicolumn{2}{|c|}{} &for one AD (\%) &Baseline &Goal \\
\hline \hline
\multicolumn{2}{|l|}{Number of protons} &0.8 &0.3 &0.1 \\
\hline
\multirow{6}{*}{Detector efficiency} &Energy cuts &0.8 &0.2 &0.1  \\
\cline{2-5}
&Time cuts &0.4 &0.1 &0.03  \\ \cline{2-5}
&H/Gd ratio &1.0 &0.1 &0.1  \\ \cline{2-5}
&n multiplicity &0.5 &0.05 &0.05  \\ \cline{2-5}
&Trigger &0.01 &0.01 &0.01  \\ \cline{2-5}
&Live time &$<$0.01 &$<$0.01 &$<$0.01  \\ \hline
\multicolumn{2}{|l|}{Total uncertainty} &1.7 &0.38 &0.18 \\
\hline
\end{tabular}
\caption{Breakdown of the Daya Bay detector systematic uncertainties.  From~\cite{DayaBay:2007}}
\label{tab:detector-sys}
\end{table}

This paper focuses specifically on the Daya Bay acrylic vessels (AVs) and their related overflow tank systems, discussing design, fabrication, transport, installation, and ongoing characterization.  The properties and history of the acrylic vessels must be particularly well-known because of the AVs' impact on the largest detector systematics: the acrylic's proximity to the target volume and effect on light transmission effects detection efficiency, while the vessels' shapes determine the size and shape of the AD's target.  In addition, reliable AVs are necessary to ensure the proper functionality of the experiment over the minimum designed experimental lifetime of 5~years.  The AV overflow tank system, which holds excesses of the liquids inside the AVs, is crucial in monitoring the mass of the liquids inside the AD to $<$0.1\%, and thus must also be thoroughly characterized.

\begin{table}[htpb]
\centering
\begin{tabular}{|c|c|c|c|}
\hline 
\ Stage &\ OAV Date &\ IAV Date  & \ Overflow Tank Date  \\ \hline \hline
Begin Construction &Dec 2008 & May 2009 & Dec 2009 \\ \hline
End Construction &May 2009 & Oct 2009 & May 2010 \\ \hline
Begin Shipping & Jun 2009 & Oct 2009 & May 2010 \\ \hline
Delivered to Daya Bay & Jul 2009 & Oct 2009 & Jun 2010 \\ \hline
\multirow{2}{*}{On-site Cleaning} & OAV1: Nov 2009 & IAV1: Nov 2009 & \multirow{2}{*}{Jul 2010} \\
 & OAV2: Jan 2010 & IAV2: Mar 2010 & \\ \hline
\multirow{2}{*}{Installed} & OAV1: Nov 2009  & IAV1: Apr 2010 & AD1: Jul 2010 \\
 & OAV2: May 2010 & IAV2: May 2010 &  AD2: Sep 2010 \\ \hline
Filled with N$_2$ Gas & Oct 2010 & Oct 2010 & Oct 2010 \\ \hline
Fill with Liquid & Spring 2011 & Spring 2011 & Spring 2011 \\ \hline
\end{tabular}
\caption{Fabrication and assembly timelines of the inner and outer acrylic vessels and overflow tank system for the first two ADs.}
\label{tab:timeline}
\end{table}

Fabrication, construction, and characterization of acrylic vessels for the first two ADs took place between late 2008 and the middle of 2010.  The timeline of the AV and overflow tank life cycles can be seen in Table~\ref{tab:timeline}.  Design and R\&D began in 2006.

\section{Vessel and Overflow Tank Design}

\begin{figure}[htb]
\centering
\includegraphics[trim=11cm 3cm 3cm 3cm, clip=true, width=0.75\textwidth]{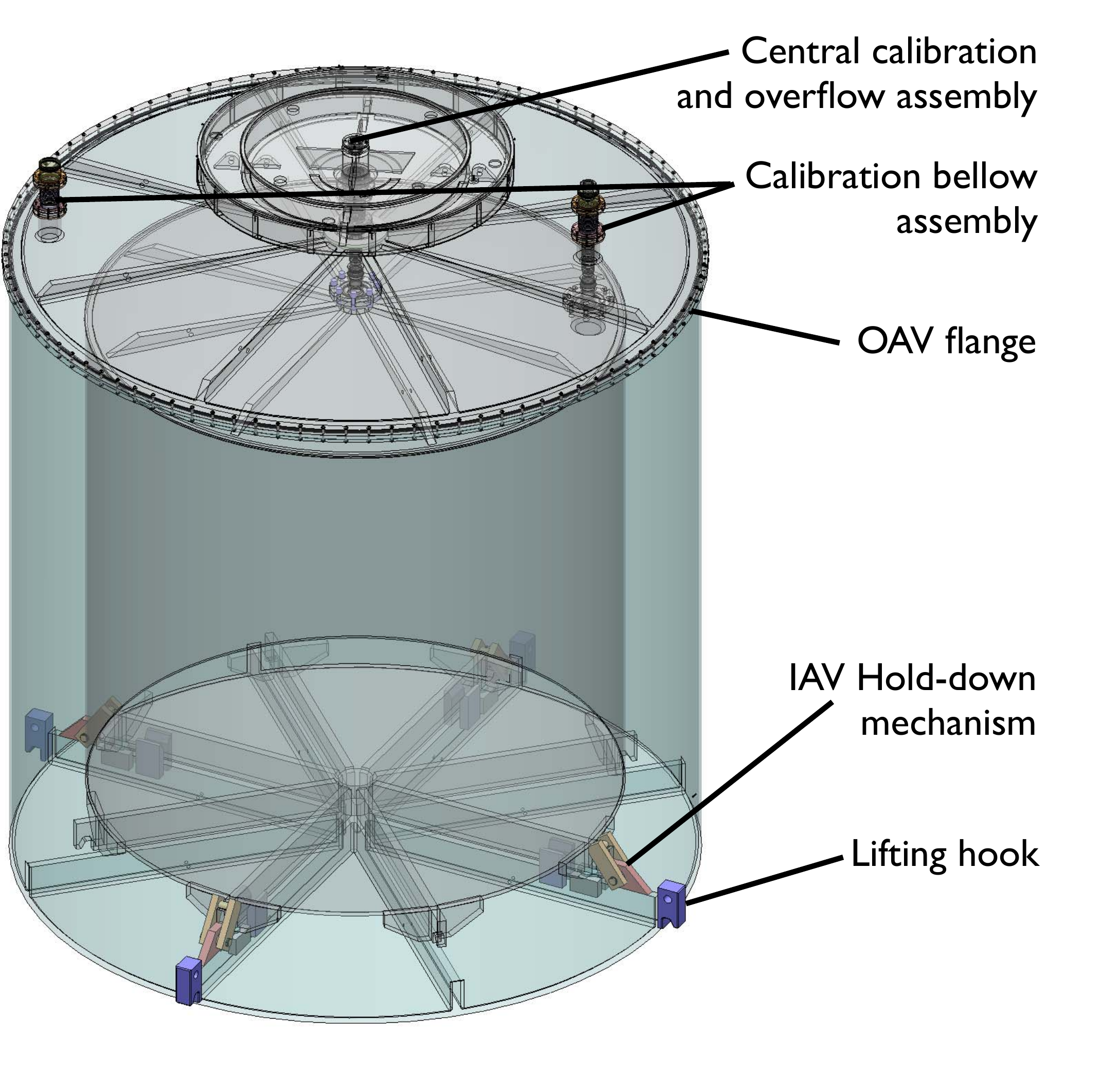}
\caption{Transparent model of the complete OAV, IAV, and overflow tank system.  The IAV is nested inside the OAV, with the overflow tank appearing above both AVs.  The overflow tanks are outside the main detector volume and above the steel lid.  The mineral oil overflow tanks are part of the stainless lid and not shown here.}
\label{fig:3m4mOF}
\end{figure}

The Daya Bay acrylic target vessel system is composed of three main elements: the inner acrylic vessel (IAV), the outer acrylic vessel (OAV), and the overflow tank system.  A complete acrylic vessel (AV) system is pictured in Figure~\ref{fig:3m4mOF}.  The IAV contains the Gd-LS volume, and is visible in Figure~\ref{fig:3m4mOF} as the inner nested cylinder.  The OAV holds LS, and is visible in Figure~\ref{fig:3m4mOF} as the outer cylinder.  The overflow tank system rests on top of the stainless steel vessel (SSV) that encapsulates the AVs, PMTs, reflectors, and AD liquids.  This subsystem contains extra separated LS and Gd-LS volumes to ensure that both AVs are completely filled at all times during running despite environmental temperature and pressure changes.

The AV structures must satisfy a wide range of design requirements.  The main purpose of the the AVs is to maintain the target and gamma-catcher liquid regions as separate nested volumes of well-defined size.  Thus, the design of the OAV must accommodate the nesting of the IAV within it.  In addition, the vessels must be structurally sound while filled or unfilled and leak-tight. The vessels must always be completely full, with any excess liquid volume closely monitored in overflow tanks.  There must also be a leak-tight liquid pathway between the vessels and the overflow tank system.  To maintain design simplicity, these pathways should also accommodate the processes of filling and calibrating the liquid regions inside the vessels.  The position of the two vessels with respect to one another should also be consistent to maintain these pathways and the liquid regions' shapes.

The proper physics response of the AD places other design constraints on the vessels.  They must be sufficiently optically clear such that anti-neutrino signals from the center regions can be detected by PMTs in the outer regions of the detectors.  They must also have low radioactivity so as not to contribute significant background to the experiment.

Further design requirements are imposed by the shipping and assembly processes.  The vessels' design must include a method of being lifted during transportation and installation.  They should also be designed to maintain their shape and integrity during and after experiencing stresses during these processes.  The design should additionally allow for all parts to fit together properly in the presence of minor misalignments or dimensional flaws.

Finally, the vessels must be designed to ensure that the AD will function properly for the entire planned length of the experiment.  This means that the materials used for the acrylic vessels and overflow tanks must be compatible with LS, Gd-LS, and mineral oil over long time periods.  The materials compatibility issues are also important for the stability of the Gd-LS.  Extensive testing and special care has been taken to ensure that the wetted materials that come in contact with the detector liquids meet these requirements.

\subsection{Vessel Design}

Figure~\ref{fig:3m4mOF} provides an overview of the acrylic vessel design.  The IAV and OAV are composed of a main cylindrical section approximately 3~m and 4~m in diameter, and 3~m and 4~m in height, respectively, and a conical top of 3\% gradient.  The walls are kept thin, 10~mm for the IAV and 18~mm for the OAV, to reduce the amount of non-scintillating absorptive material in the inner detector.  The vessels encapsulate a volume of 23 tons of Gd-LS in the target and 25 tons of LS in the gamma-catcher region.  The vessels are manufactured from ultraviolet-transmitting (UVT) acrylic, which allows for maximal transmission of near-UV and optical scintillation light emitted by the AD liquids.  Acrylic is the vessel material of choice because it is stronger, cheaper, lower in radioactivity, and easier to fabricate than most other optical window materials.    Fabrication and construction of large acrylic objects is a common practice: acrylic is the material of choice for commercial aquariums and other applications, and was also used to build the 12~m diameter sphere inside the Sudbury Neutrino Observatory~\cite{SNODET}.

\begin{figure}[htb]
\centering
\subfigure[Drawing of the OAV flange double o-ring seal, including leak check port.] {
  \includegraphics[width=0.45\textwidth]{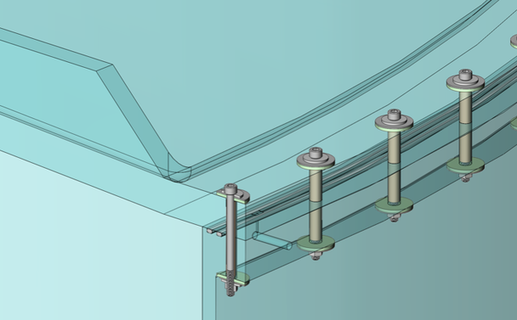}
  \label{subfig:4mSeal}
}
\hspace{0.5cm}
\subfigure[Photograph of the as-built OAV flange-lid seal, including leak check port and plug.] {
  \includegraphics[width=0.45\textwidth]{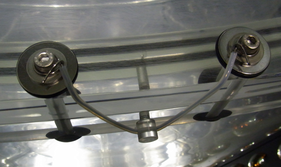}
  \label{subfig:OAVFlangeReal}
}
\caption{Close-up drawing and photograph of the OAV flange double o-ring seal connecting the OAV lid to the rest of the OAV vessel.  Also pictured in the photograph is the OAV leak-check port and the plug used to seal the leak-check port and the space between o-ring grooves.   A teflon-covered stainless wire is used to hold the port plug in place. }
\label{fig:4mSeal}
\end{figure}

The nested design of the inner and outer acrylic vessels in Daya Bay poses an interesting challenge.  The OAV is manufactured in the US and shipped to Daya Bay, while the IAV is fabricated in Taiwan. The vessels are assembled together into the AD in the Surface Assembly Building (SAB) at Daya Bay.  To accommodate this fabrication plan, the conical top of the OAV has been designed as a detachable lid that connects to the rest of the vessel via a double o-ring seal to a 4~m-diameter flange on top of the OAV walls.  A cross-section of this OAV flange-lid connection can be seen in Figure~\ref{fig:4mSeal}, along with a photo of a production OAV flange seal and accompanying leak check port plug.  The o-rings are compressed by stainless bolts torqued into nitrogen-allowed nuts at 60~foot-pounds.  The bolts are separated from the acrylic by viton washers and teflon bolt-hole sleeves.

\begin{figure}[t]
\centering
\subfigure[Cutaway of the connection between the IAV calibration port and the IAV calibration box.] {
  \includegraphics[width=0.4\textwidth]{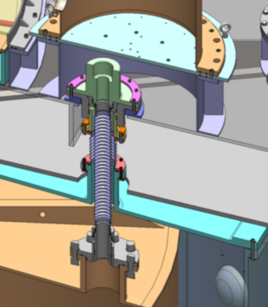}
  \label{subfig:IAVCalPort}
}
\hspace{0.1cm}
\subfigure[Cutaway of the connection between the OAV calibration port and the OAV calibration box.] {
  \includegraphics[width=0.47\textwidth]{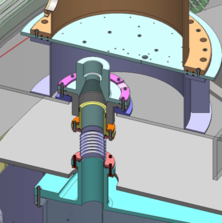}
  \label{subfig:OAVCalPort}
}
\caption{Detailed close-up of connections between the off-center AV calibration ports and connection hardware on the top of the stainless steel tank lid.  Parts are color-coded to help differentiate between individual pieces.  O-ring locations are also pictured.}
\label{fig:OffCenter}
\end{figure}

The top of the IAV and the OAV lid have two and three ports, respectively, which will serve as the entry point for liquids, and entry and exit points for calibration sources.  A close-up view of the off-center calibration ports can be seen in Figure~\ref{fig:OffCenter}.  The off-center ports accommodate delivery of LEDs and radioactive sources during calibration and liquids during filling, while the central port allows delivery of calibration sources and allows the flow of liquids between the AVs and the LS and Gd-LS overflow tanks.  The IAV ports are connected to semi-flexible teflon bellows that run up to connection hardware on the stainless steel tank lid.  While traversing the OAV region, the IAV teflon bellow runs inside a larger-diameter teflon bellow that connects the OAV ports to a related set of connection hardware on the stainless steel vessel lid.  The calibration sources are housed in and lowered from slightly elevated automated calibration units directly above the stainless steel vessel connection hardware.  All the connecting parts are made of acrylic, teflon, or viton, and utilize single or double o-ring seals to prevent liquid leakage.  To allow for proper connection despite minor radial and height misalignments, the calibration tubes are capable of sliding up and down with respect to other parts while maintaining leak-tightness.  The flexibility of the teflon bellows also helps assure connection in spite of slight misalignments.

\begin{figure}[htpb]
\centering
\subfigure[Results of FEA calculation.  The small region of maximum stress at 9.5~MPa is labelled ``MX''.  Another high-stress region can be seen at the bottom middle of the OAV.] {
  \includegraphics[width=0.46\textwidth]{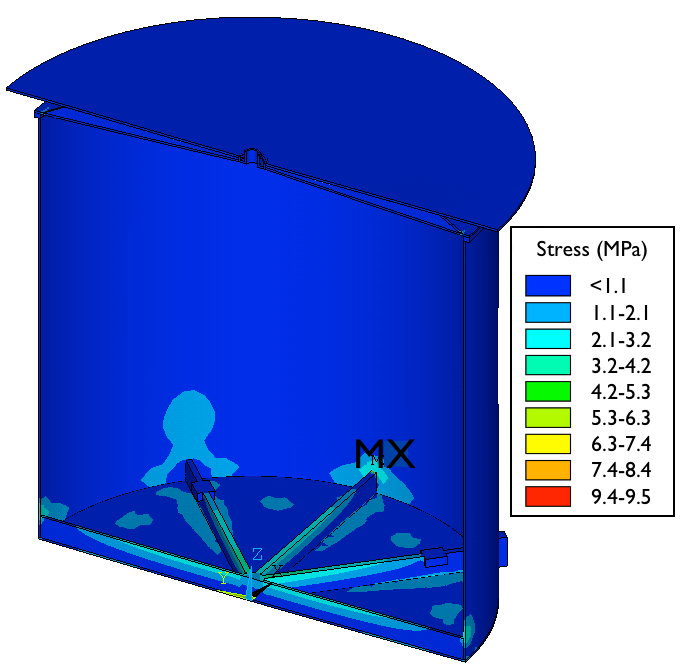}
  \label{subfig:FEACalc}
}
\hspace{0.1cm}
\subfigure[A depiction of the filling scenario giving rise to the stresses calculated via FEA in the adjoining figure.] {
  \includegraphics[trim=6.3cm 2cm 8.6cm 3cm, clip=true, width=0.4\textwidth]{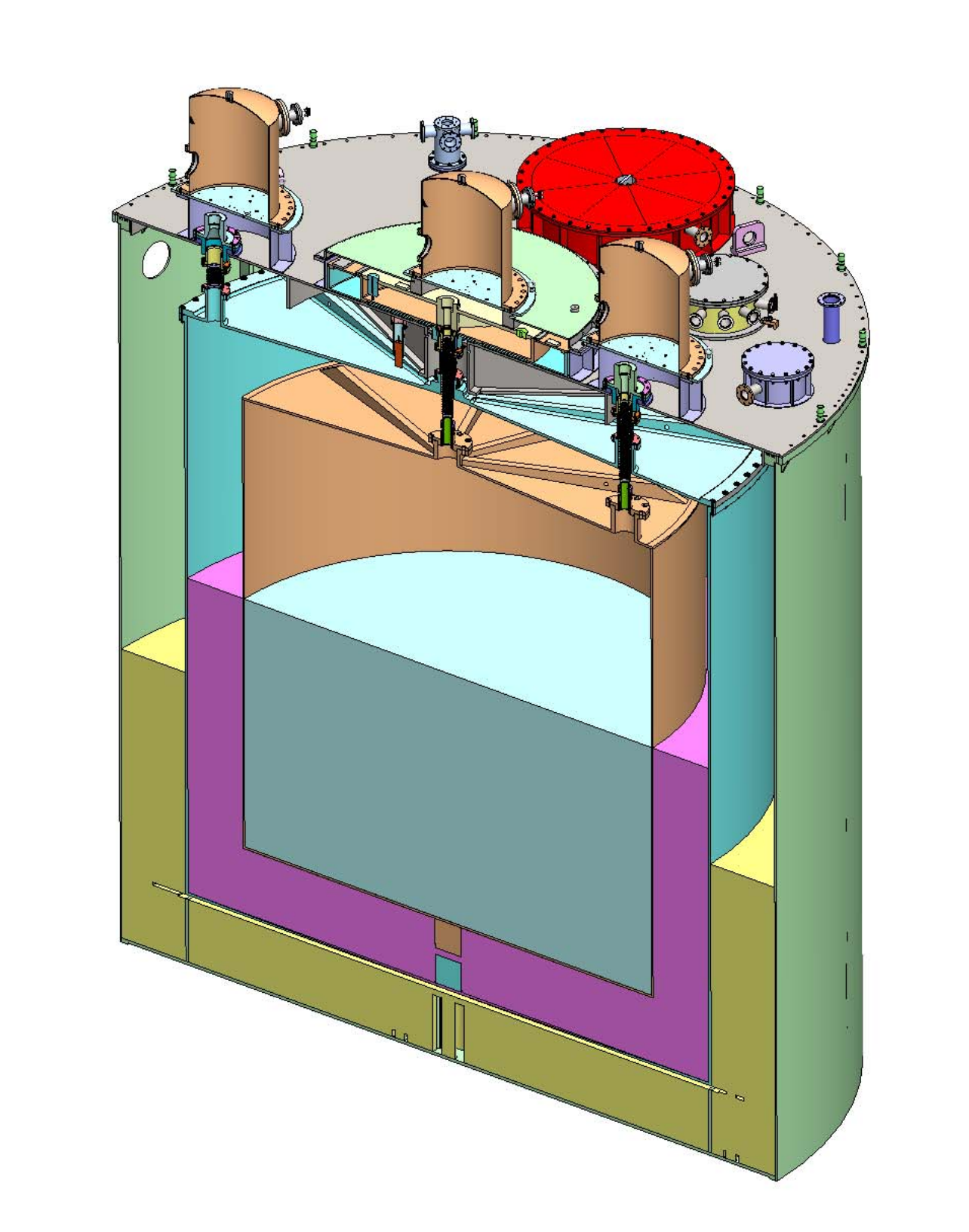}
  \label{subfig:FEAFig}
}
\caption{FEA results for a scenario in which the liquid level is 30~cm higher on the inside of the OAV than on the inside.}
\label{fig:FEA}
\end{figure}

The bottoms of the AVs provide structural support, facilitate the alignment of the acrylic vessels, and make a mechanical link between the inner and outer vessels through the hold-down mechanisms.  5~cm-thick supporting ribs can be seen in Figure~\ref{fig:3m4mOF} on the bottom outside of the IAV and the bottom inside of the OAV, as well as on the outside of the IAV and OAV lids.  The ribs provide a support structure that allows the vessels to withstand stresses experienced during lifting, transport, and filling.  Finite element analysis (FEA) was carried out on the vessel designs to determine the necessary wall thickness and structural support.  The highest stress risk occurs during the filling process if liquid levels become uneven: Figure~\ref{fig:FEA} shows that expected stresses for a 30~cm level difference inside and outside the OAV are as high as 10~MPa.   To avoid these stresses, liquid levels will be matched to 5~cm during filling.

The long-term design stress limits for the acrylic vessels is 5~MPa.  Short-term stress exposures up to 10~MPa are considered safe.  The filling process, which takes 4 days to complete, is considered a short duration compared to the nominal 5-10 year lifetime of the acrylic vessels.  Stress characterization tests that validate these tolerances are discussed in Section 6.1.4.

\begin{figure}[htpb]
\centering
\subfigure[A close-up of the lateral alignment guide, support puck, IAV hold-down mechanism, rib stop, and lifting hook without an IAV present.] {
  \includegraphics[width=0.45\textwidth]{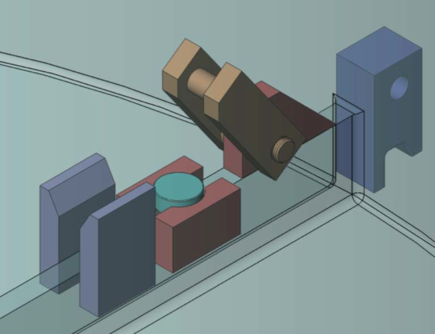}
  \label{subfig:HoldDownEmpty}
}
\hspace{0.5cm}
\subfigure[A close-up of the as-built components pictured to the right.  Not present in this picture is the support puck and IAV hold-down latch.] {
  \includegraphics[width=0.45\textwidth]{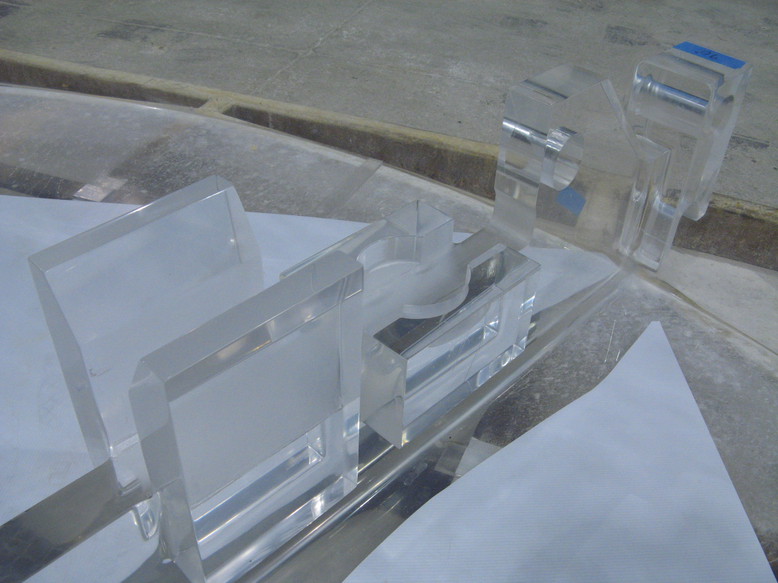}
  \label{subfig:HoldDownReal}
}
\hspace{0.5cm}
\subfigure[A close-up of an engaged IAV hold-down mechanism.  IAV and OAV lifting hooks are also visible, along with the engaged lateral alignment guide.] {
  \includegraphics[width=0.5\textwidth]{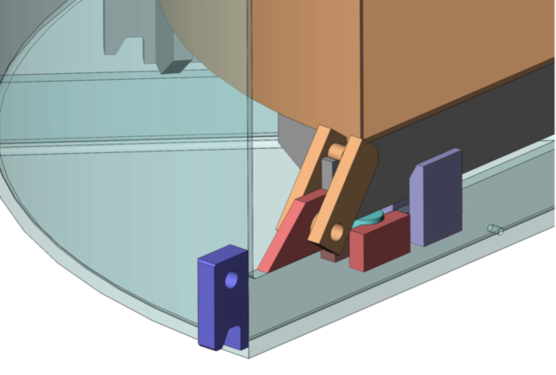}
  \label{subfig:HoldDown3m}
}
\caption{Close-up views of the IAV/OAV interface.  The IAV rests on the teflon puck (blue), which is shimmed beforehand to ensure IAV levelness, The latch (brown) is rotated to secure the IAV with respect to the OAV.  The lateral alignment guide directs azimuthal placement while the IAV is being lowered onto the shimmed puck.}
\label{fig:HoldDown}
\end{figure}

A few other design features located at the bottom of the vessels can be seen in Figure~\ref{fig:HoldDown}.  Lifting hooks on the IAV and OAV can be seen in Figure~\ref{subfig:HoldDown3m}; four hooks are located on the outside of each AV, one every 90$^{\circ}$.  Rigging and lifting of the vessels using these hooks will be discussed further in Section~5.  Figure~\ref{fig:HoldDown} also demonstrates the hold-down mechanism which latches the IAV to the OAV.  The IAV is lowered into the vessel and four of its ribs are rested on four teflon pucks that rest on the ribs of the OAV.  A latching mechanism connected to the OAV is then rotated into place over a hook on the end of the IAV rib.  This mechanism is designed to constrain the IAV's movement upward.  The latch can only be disengaged by applying upward force to the bottom of the latch, minimizing the risk of accidental disengagement from vibration during transport.  The rib stops atop the OAV ribs beyond the IAV rib ends constrain the IAV's movement in the radial direction.  One set of rotational stops attached to either side of one OAV rib, along with the rib stops, constrain motion about the axis of the vessel.

\begin{table}[htp]
\centering
\begin{tabular}{|c|c|c|c|}
\hline
\multirow{2}{*}{Material} & \multicolumn{3}{|c|}{Compatible With:} \\ \cline{2-4}
& LS & Gd-LS & MO  \\ \hline \hline
Acrylic & Y & Y &Y  \\ \hline
Teflon & Y & Y&Y \\ \hline
Viton & Y & Y&Y  \\ \hline
Nitrogen Gas & Y & Y&Y  \\ \hline
Stainless Steel & N & N &Y \\ \hline
\end{tabular}
\caption{Material compatibility between AV and overflow system construction materials and AD liquids.}
\label{tab:compatibility}
\end{table}

Table~\ref{tab:compatibility} lists all the materials present in the AVs, overflow tanks, and connection hardware, and overviews the level of compatibility between these materials and the AD liquids.  Materials are non-compatible if one causes the mechanical or optical degradation of the other when the two are in contact.  Further documentation of compatibility QA measurements are detailed in Section 6.1.3.  The vessel and port connections only use materials compatible with all AD liquids.

\subsection{Overflow Tank Design}

\begin{figure}[htpb]
\centering
\subfigure[Drawing of overflow tank and connection hardware.] {
  \includegraphics[trim=6cm 8cm 8cm 5cm, clip=true, width=0.5\textwidth]{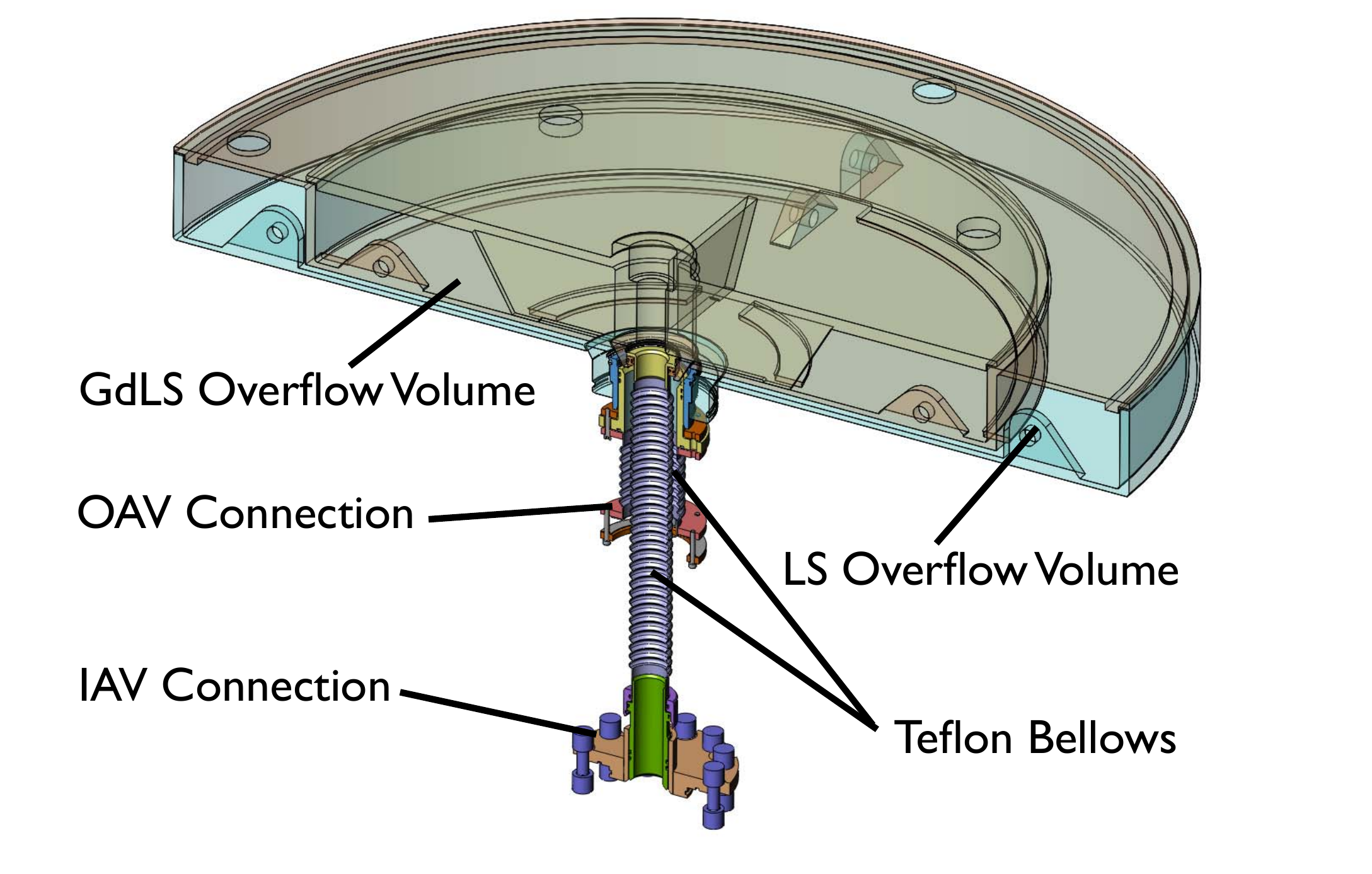}
  \label{subfig:OverflowClose}
}
\hspace{0.5cm}
\subfigure[Photograph of an as-built overflow tank.  The two overflow regions are clearly visible.] {
  \includegraphics[width=0.4\textwidth]{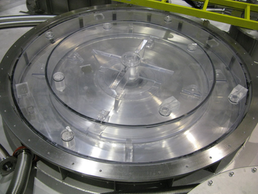}
  \label{subfig:OverflowInstall}
}
\caption{Close-up view and photograph of the overflow tank system, including OAV/IAV connection hardware.  The parts are color-coded to emphasize the individual pieces.  The stainless steel overflow tank cover and spider ring are not pictured in either drawing or photograph.}
\label{fig:Overflow}
\end{figure}

Overflow tanks are connected to the central detector volumes for Gd-LS, LS, and MO to allow for the thermal expansion of the detector liquids during filling, transport, and storage.  A close-up view of the overflow tanks can be seen in Figure~\ref{fig:Overflow}.  The overflow tanks consists of two separated nested spaces bounded by acrylic. The innermost region, the Gd-LS overflow tank, is  1.3~m wide and 13~cm deep, and is surrounded by an LS overflow tank of 1.8~m diameter and 13~cm depth.  This corresponds to an overflow liquid volume of 167~liters Gd-LS and  151~liters LS.

These overflow spaces are created by three large acrylic structures, which can be see in Figure~\ref{fig:Overflow}. The first is a 0.9~m-radius flat-bottom cylindrical tank with no top that serves as the base of the LS overflow region.  The next is a 0.65~m-radius cylinder with an extra elevated flared ring of material on its outside top that rests on top of the outer cylinder.  Raised areas on the underside of this piece allow it to rest on the bottom of the outermost cylinder, while still allowing liquid to pass below it inside the outermost cylinder.  This second piece is the base of the Gd-LS overflow region and the flared covers the LS overflow region on top.  The third piece is a lid for the Gd-LS overflow region.  Polyurethane is applied to the interfaces between these three pieces to provide a seal between the overflow tank volumes and the outer cover gas region of the detectors.  A stainless steel ring with spokes, or spider ring, is fastened down on top of these acrylic pieces to hold them together without the need of screws in acrylic.  Finally, a stainless steel shell surrounding the apparatus provides an air- and liquid-tight seal to enclose the nitrogen gas and AD liquids in the overflow tanks and separate them from the water outside the detector.

The AD is filled with liquid by pumping Gd-LS and LS in through the off-center IAV and OAV ports, respectively, displacing nitrogen gas that is circulated through the AD before filling.  The liquid fills up through the volume of the detector, then through the bellows and into the overflow tanks.  Filling was stopped with the liquid levels at 4.6~cm and 6.4~cm above the bottom of the Gd-LS and LS overflow tanks, respectively.  An approximate 3$^{\circ}$C rise in temperature and subsequent expansion of the AD liquids would cause the complete filling the LS and Gd-LS overflow tanks.  In addition, if the temperature is lowered by 2$^{\circ}$C, the liquid will contract out of the overflow tank and begin emptying the target volume, causing unknown changes in optical properties at the top of the detector.  Because of this, a 2$^{\circ}$C temperature change is the operational limit that can be experienced by the AD during filling and operation of the detectors.

%

\begin{figure}[htb]
\centering
\includegraphics[trim=6cm 16cm 3cm 2cm, clip=true, width=5.5 in]{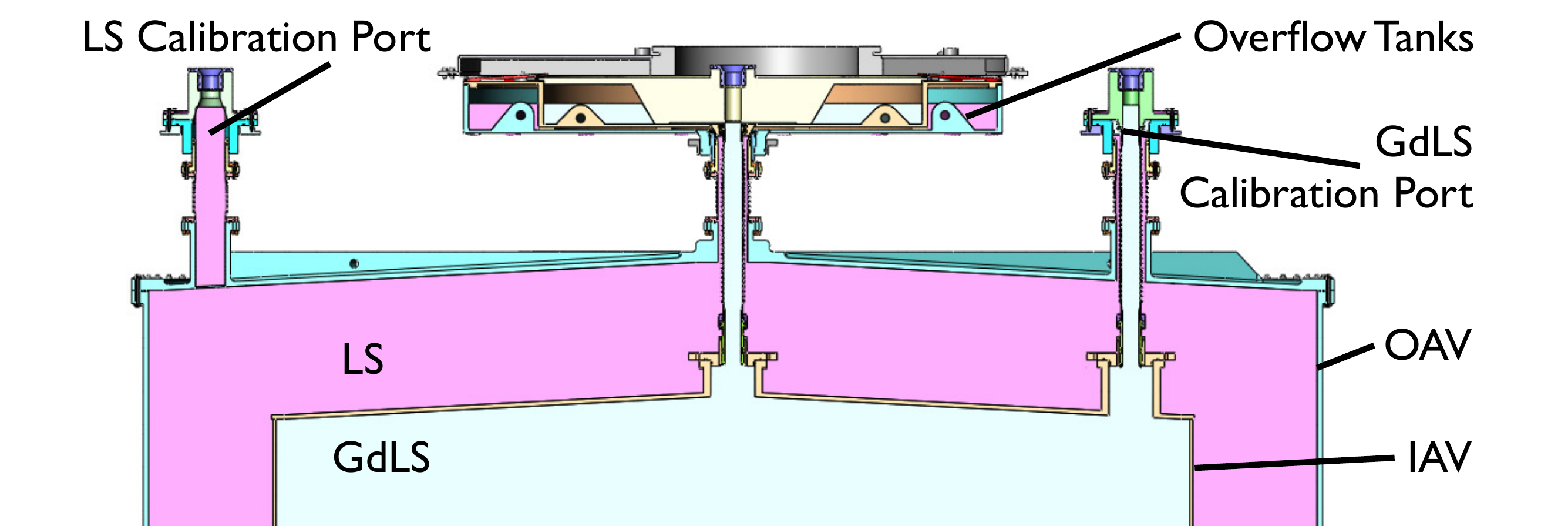}
\caption{Cross-section of a filled AD.  The path from the LS and Gd-LS bellows through to the overflow tanks are visible in this figure.  Also note the LS and Gd-LS filling up their respective calibration ports.}
\label{fig:Filled}
\end{figure}

The locations of LS and Gd-LS liquid levels in the AVs and overflow tanks for a filled AD are illustrated in Figure~\ref{fig:Filled}.  During detector operation, the liquid level is monitored in the overflow tanks by ultrasonic and capacitance sensors and in the off-center calibration ports with cameras to determine changes in the target and gamma-catcher mass resulting from changes in temperature and density the liquids.


\section{Fabrication}

\subsection{OAV Fabrication}

\begin{figure}[htpb]
\centering
\subfigure[Bond locations on the OAV barrel.  Top cylinder bonds are dashed, bottom cylinder bonds are dotted, and circumferential bonds are solid.] {
  \includegraphics[width=0.4\textwidth]{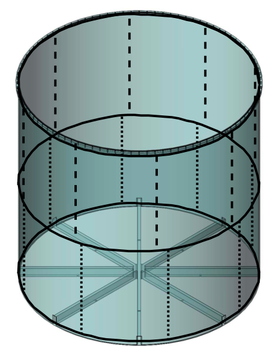}
  \label{subfig:oavbonds1}
}
\hspace{0.5cm}
\subfigure[Bond locations on the OAV bottom.  Solid lines indicate bonds connecting the ribs and hub to the main vessel.  Dotted lines indicate the bottom-barrel bond and the bond joining the two bottom sheets.  All colored parts are also bonded onto the main structure.] {
  \includegraphics[width=0.45\textwidth]{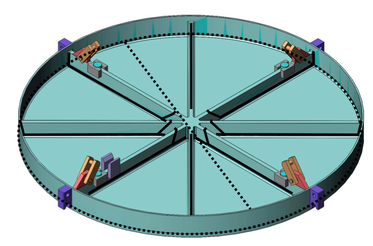}
  \label{subfig:oavbonds2}
}
\hspace{0.5cm}
\subfigure[Bond locations on the OAV top: directly through the lid middle and around the OAV lid ports.] {
  \includegraphics[width=0.3\textwidth]{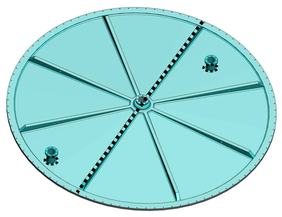}
  \label{subfig:oavbonds3}
}
\caption{Approximate bond locations on the OAVs.}
\label{fig:oavbonds}
\end{figure}

OAVs are manufactured at Reynolds Polymer Technology, Inc., in Grand Junction, Colorado, USA~\cite{Reynolds}.  The walls of the OAVs are constructed from 16 sheets of Polycast UVT acrylic~\cite{Polycast}.  The bottom, lid, and flange are cut from two large, thick blocks of Reynolds-cast UVT acrylic.  The bottom ribs are made from 50~mm-thick sheets of Reynolds UVT acrylic and a central hub piece machined from a block of Reynolds UVT acrylic.  The vessel elements were bonded using a proprietary UVT bonding syrup from Reynolds; the bond material appears in approximately 1/8'' wide regions between bonded sheets.  A drawing showing the placement of acrylic sheets and bond lines on an OAV can be seen in Figure~\ref{fig:oavbonds}.

OAV fabrication was preceded by the construction of a prototype OAV at the same facility, as well as a round of prototype acrylic optical testing to determine an acceptable acrylic for the production OAVs.  Radioactivity, compatibility, stress, and optical testing results for a range of acrylics are presented Section 6.1.

The cylindrical OAV walls are composed of two 4~m-diameter acrylic cylinders, each created by bending and bonding together eight acrylic sheets.  The OAV top, bottom, and flange are machined out from two thick blocks of bonded UVT acrylic.  Rib sections and a central hub are bonded to the bottom, dividing it into octants.  Additional acrylic pieces for the IAV support structure and hold-down mechanisms are bonded to four of the ribs.

Annealing is done on the individual lid, wall, and bottom components to cure bonding material and to clear residual stresses introduced during fabrication.  Care was taken to fully support the shape of acrylic components during annealing to avoid permanent deformation from components becoming more plastic at high annealing temperatures and sagging under their own weight.  Such sagging was experienced during annealing of the OAV lid prototype.  Annealing of subsequent OAV lids was done on a conical frame, while annealing of the OAV bottoms was done on precision-cut flat surfaces to ensure flatness of the OAV base and flange.  

Bonding of OAV components is done using proprietary bonding syrup mixed by Reynods.  After setting dams along the intended bond against the joining acrylic pieces, syrup is injected between the pieces and cured using a proprietary heating regimen.  The OAV flange and bottom structure are bonded to the top and bottom cylinders, respectively.  The top and bottom cylinders are then bonded together to complete the vessel construction.  A final anneal is done after bonding to clear residual stresses introduced during bonding.

\begin{figure}[htpb]
\centering
\subfigure[Image of fabricated OAV top after cleaning.  Notice the extra calibration port for the LS region.] {
  \includegraphics[width=0.45\textwidth]{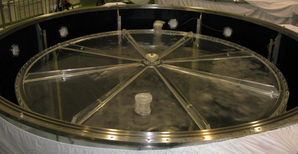}
  \label{subfig:OAVTop}
}
\hspace{0.5cm}
\subfigure[Image of fabricated IAV top during installation.  Notice in particular the bonded lid, unlike the OAV. ] {
  \includegraphics[width=0.45\textwidth]{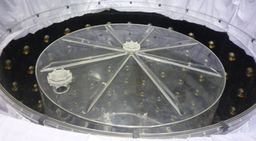}
  \label{subfig:IAVTop}
}
\hspace{1cm}
\subfigure[Image of an OAV bottom after finishing fabrication.  Notice that the ribs and extra features are on the OAV inside.] {
  \includegraphics[width=0.45\textwidth]{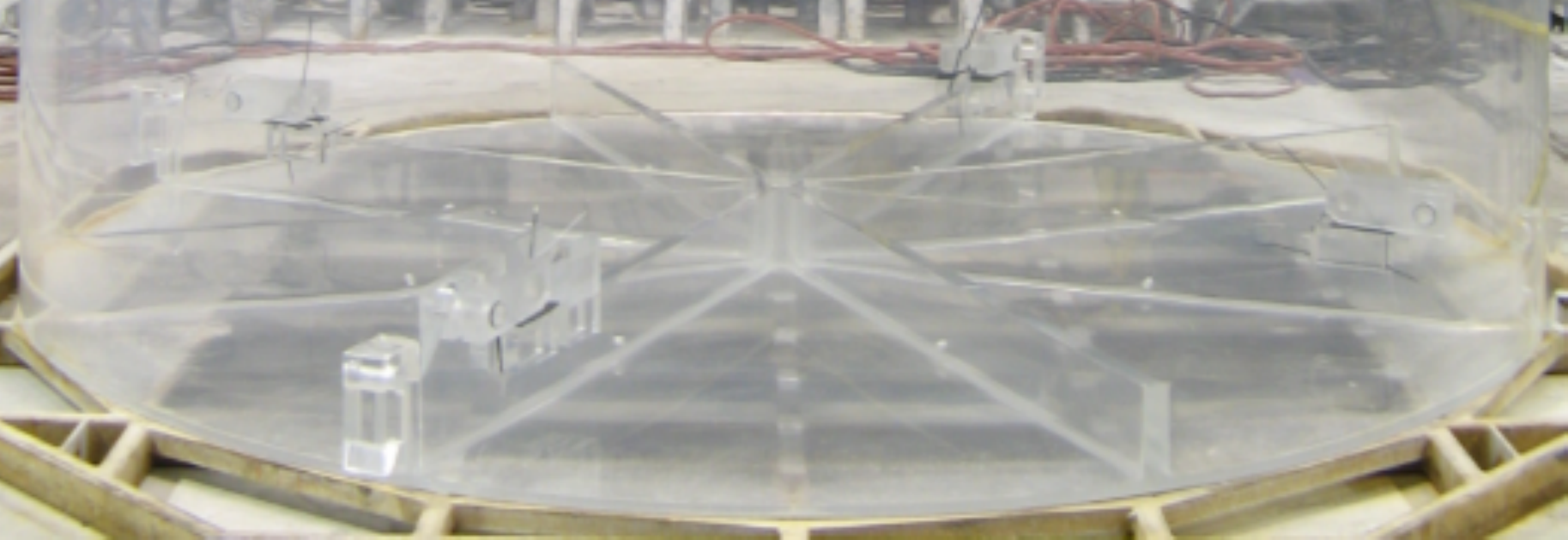}
  \label{subfig:OAVBottom}
}
\hspace{0.5cm}
\subfigure[Image of a fabricated IAV bottom during cleaning.  Notice that all structural support is located on the IAV outside.] {
  \includegraphics[width=0.45\textwidth]{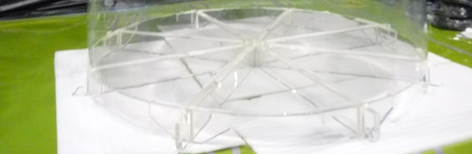}
  \label{subfig:IAVBottom}
}
\caption{A comparison of the tops and bottoms of production OAVs and IAVs.}
\label{fig:OAVIAV}
\end{figure}

Vessel surfaces and bond lines are sanded and polished after bonding to remove excess syrup and discontinuities and to improve the clarity of the vessel surface.  The OAV and IAV are polished down to 1 to 3~micron grit using aluminum oxide polishing powder, buffing wheels and water.  A fully fabricated OAV's top and bottom can be seen in Figure~\ref{fig:OAVIAV}.

\subsection{IAV Fabrication}

\begin{figure}[htpb]
\centering
\subfigure[Bond locations on the IAV barrel.  Vertical bonds are solid and circumferential bonds are dotted.  The top circumferential bond connects the IAV barrel to the IAV top, which is not pictured.] {
  \includegraphics[width=0.4\textwidth]{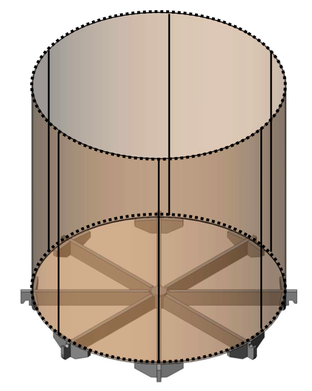}
  \label{subfig:iavbonds1}
}
\hspace{0.5cm}
\subfigure[Bond locations on the IAV bottom.  Dotted lines indicate bonds connecting the ribs and hub to the main vessel.  Dashed lines indicate the bonds joining the four main bottom sheets.  The hub is also bonded to the IAV bottom.] {
  \includegraphics[width=0.5\textwidth]{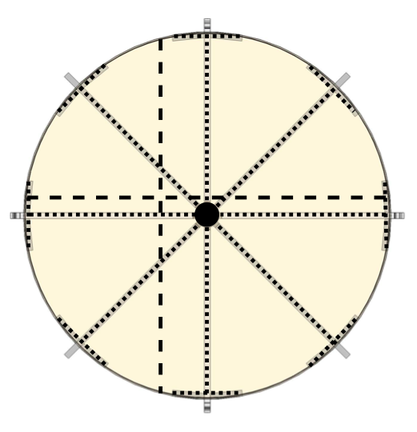}
  \label{subfig:iavbonds2}
}
\hspace{0.5cm}
\subfigure[Bond locations on the OAV top.  Bond line indications are similar to that of the IAV bottom, with the addition of solid lines indicating bond connecting the IAV top ports.] {
  \includegraphics[width=0.3\textwidth]{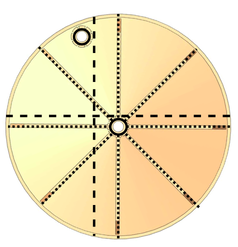}
  \label{subfig:iavbonds3}
}
\caption{Approximate bond locations on the IAVs.}
\label{fig:iavbonds}
\end{figure}

IAVs were constructed at Nakano International Limited, in Taipei, Taiwan~\cite{Nakano}.  Nakano used PoSiang UVT acrylic for all parts: six 10-mm thick sheets for the walls, four 15-mm thick sheets for the top lid, four 15-mm sheets for the bottom lid, two 55-mm sheets for the ribs, and two UVT acrylic calibration ports.  Sheets are joined together by approximately 1/8'' wide bonding regions made out of a cured UVT syrup.  A drawing showing the placement of acrylic sheets and bond lines on an IAV can be seen in Figure~\ref{fig:iavbonds}.

As with the OAVs, quality assurance measurements are done on candidate IAV materials to ensure the suitability of the production acrylic for the Daya Bay detector.  Details of these measurements are discussed in Section~6.  In addition, a prototype IAV was fabricated to test the construction processes and vessel design.

Vessel construction begins by forming the six wall sheets and then bonding them together using the UVT syrup and a system of dams to keep the syrup in place while the bonds undergo UV-curing.  During this process, the top and bottom of the vessels are constructed separately in the same manner, with the top being baked on a mold to achieve a conical shape.  In addition, ribs are bonded onto the IAV bottom and top, and center and off-center calibration ports are bonded to the IAV top.  Following construction, each separate component's surfaces are polished in a similar manner as described in the OAV construction.  After polishing, the top is bonded to the walls.  The inside of the vessel must then be cleaned before the final bond is made, as after this point no further access to the inside of the IAV exists.  The cleaning of the IAV will be described in greater detail in Section~5.1.  After cleaning, the bottom is bonded to the rest of the vessel, completing construction.  The IAV bottom edge must be properly shaped to encourage excess bonding syrup to flow out of the vessel during curing rather than into the vessel where it cannot be removed or polished out.  A picture of a fabricated IAV's top and bottom is visible in Figure~\ref{fig:OAVIAV}.

In Figure~\ref{fig:OAVIAV} one can identify contrasting features on the IAV and OAV top and bottom .  On the top, the major differences are the presence of the extra OAV flange seal and an extra calibration port.  On the bottoms, the major difference is that the OAV's support structure is inside the OAV, while the IAV's support structure is on the outside of the IAV.  

\subsection{Overflow Tank Fabrication}

The three main acrylic components for the overflow tanks are the LS overflow tank, the Gd-LS overflow tank, and the tank lid.  The acrylic pieces that comprise each of these three components are fabricated by subcontractors of Reynolds by cutting and forming sheets of UV-absorbing (UVA) acrylic into designed shapes.  The component sub-pieces are then sent back to Reynolds to be bonded into the three separate overflow components using Weld-on 40 acrylic cement.  Other smaller acrylic parts from the overflow tank system and connection hardware are also manufactured by Reynolds' subcontractors.

\section{Transport and Shipping}

\begin{figure}[htb]
\centering
\includegraphics[width=6.5 in]{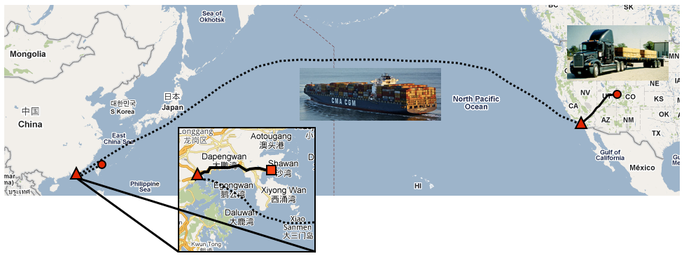}
\caption{A map of the entire shipping route for OAVs and IAVs.  Circles indicate the origin locations of OAV (Grand Junction, CO, USA) and IAV (Taipei, Taiwan), triangles indicate the ports of Yantian and Long Beach, through which the AVs travel, and the square indicates the final destination, Daya Bay, China.}
\label{fig:route}
\end{figure}

After fabrication, the IAVs and OAVs are transported from their respective locations to the Surface Assembly Building (SAB) at Daya Bay, where they are assembled into the antineutrino detectors.  The OAVs are manufactured at Reynolds Polymer Technology in Grand Junction, CO and are transported by truck to Long Beach, CA, by ship to Yantian Port, and finally by truck to Daya Bay.  The IAVs must be trucked from the Nakano factory in Taipei to port, and then shipped to Yantian, where they are then brought by truck.  The entire shipping route of the vessels can be seen in Figure~\ref{fig:route}.

\subsection{Shipping Crates}

\begin{figure}[htpb]
\centering
\subfigure[ISO drawing of an OAV in its shipping container.  The lifting frame is also included in this drawing.] {
  \includegraphics[trim=4cm 2cm 6cm 2cm, clip=true, width=0.51\textwidth]{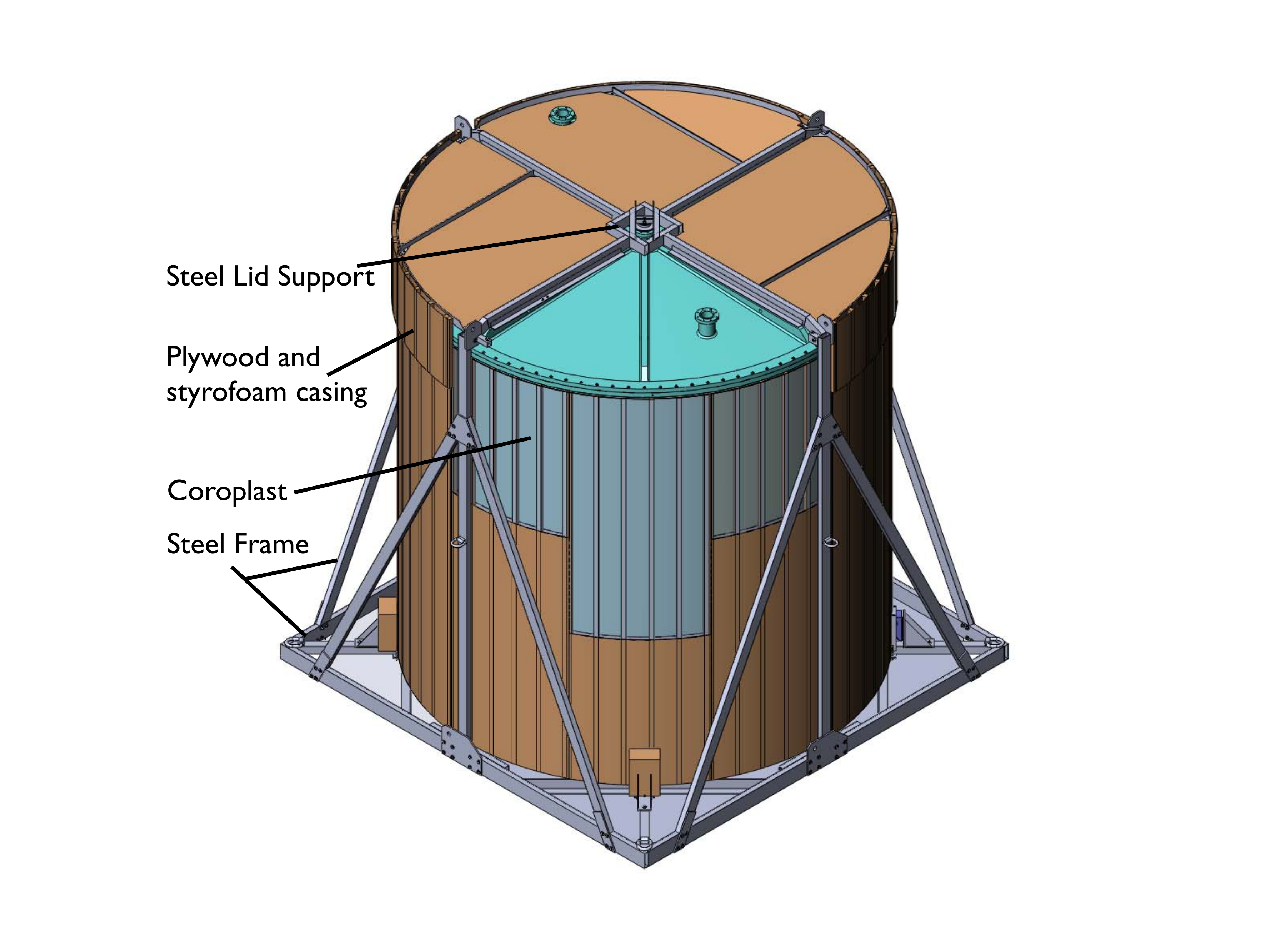}
  \label{subfig:OAVShipDraw}
}
\hspace{0.1cm}
\subfigure[A photograph of a production OAV in its shipping frame.] {
  \includegraphics[width=0.39\textwidth]{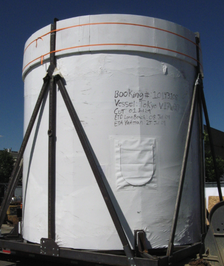}
  \label{subfig:OAVShipPhoto}
}
\caption{An isometric drawing and photograph of the OAV shipping frame.}
\label{fig:OAVShip}
\end{figure}

For transport, the vessels are packed into shipping frames and containers that support the structure of the AVs and minimize their exposure to light and environmental conditions.  The OAV shipping frame can be seen in Figure~\ref{fig:OAVShip}.  The sealed OAV rests on a steel square base and is secured by fastening hold-down pieces over all four OAV lifting hooks.  A steel A-frame structure rises on each side of the base.  A steel OAV lid support structure is lowered onto the top middle of the four A-frames and is attached to ears located there.   A plug in the central OAV port is screwed upward into the center of the lid support structure, providing extra structural support for the lid during shocks and preventing resonant vibrations in the lid during shipping.  The OAV is protected from impacts by casing made of polystyrene and kiln-treated wood and plywood.  Underneath this casing are layers of self-adhesive plastic film and Coroplast~\cite{Coroplast} to preserve the surface quality of the OAV.  To ensure that the inside of the vessel remains somewhat isolated from the surrounding environment, a hose and HEPA filter is connected to both off-center ports.

\begin{figure}[htb]
\centering
\includegraphics[width=3.0 in]{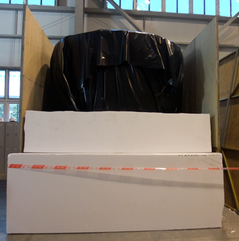}
\caption{Photograph of a partially-unpacked IAV in its container.  Some of the styrofoam packing has been removed, but the IAV remains in its black plastic casing with styrofoam in front and behind it.}
\label{fig:IAVShip}
\end{figure}

The IAV shipping container can be seen in Figure~\ref{fig:IAVShip}.  The container consists of six steel-framed panels lined on the inside with wood panels.  These panels are bolted together to form the cubical shipping crate.  Foot-thick styrofoam panels are placed inside these outer panels to support the IAV on all sides and to absorb shocks during shipping.   Lid support is less crucial for the IAVs, as they are smaller in diameter that the OAVs.  To protect its surfaces, the IAV is coated in plastic wrap and then covered with black plastic sheeting.  The IAV ports are covered with flanges containing HEPA filters to ensure that the IAV insides retain their cleanliness.

\subsection{Shipment Monitoring}

Several devices are used to monitor OAV location, acceleration, temperature, humidity, pressure and light exposure during shipment.  Customs issues in shipping prevented the application of the same monitoring regimen for the IAVs; as a result, no shipping data is available for IAV1 and IAV2.  

For the OAVs, acceleration data is collected every 9 seconds by MSR 145W dataloggers~\cite{MSR} mounted on the base of the shipping frame and the OAV lid.  Additionally, the MSRs on the base recorded temperature and pressure readings every 90 seconds.  The lid-mounted MSR had a pressure sensor on an external cable, which allowed measurement of the pressure inside the OAV.  The largest pressure changes correspond to changes in altitude as the OAVs are trucked from $\sim$1400~m to sea level.  Trackstick Pro GPS data loggers are also mounted on the shipping frame to provide location information while the OAVs are in the United States~\cite{Trackstick}.  To measure shocks, such as would occur if the AV is dropped, Shock Timer 3d loggers are mounted along with the MSRs on the shipping frame and OAV lid \cite{IST}.  These sensors recorded the date and time of all shocks over 3~g and took additional temperature and humidity readings every hour.  Finally, a pair of HOBO Pendant Dataloggers are mounted from a threaded rod suspended inside each AV to monitor light levels~\cite{HOBO}.  These are expected to be low as the OAVs are shipped with opaque coverings to shield them from UV.

\begin{figure}[htb]
\centering
\includegraphics[trim=6cm 5cm 6cm 7.5cm, clip=true, width=0.8\textwidth]{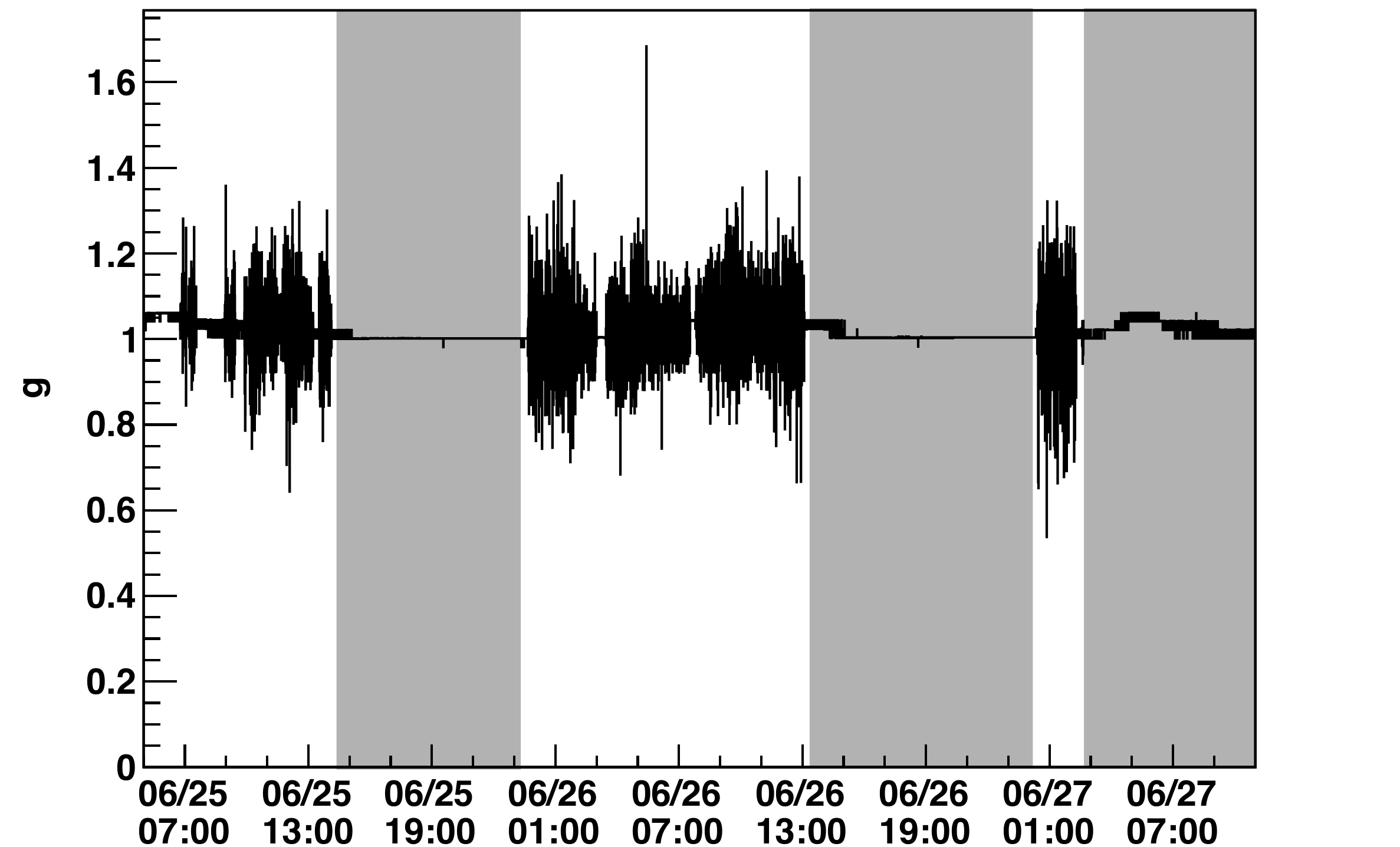}
\caption{Periodic acceleration measurements, measured in g.  The measurement period begins in Grand Junction, Colorado and ends at port in Long Beach, California.  The gray-shaded periods of low acceleration correspond to times when the OAVs and transport truck were at rest.}
\label{fig:accel}
\end{figure}

The first pair of OAVs left CO on June 25, 2009 and arrived at Daya Bay on August 5 and August 6.  Just outside of Grand Junction, metal lifting tabs attached to the top of the OAV lid support structure were bent down by impact with a low bridge; this impact was recorded by the shock timers.  The truck was returned to Reynolds to ensure that the vessels were not damaged.  Other shocks greater than 3~g were recorded 10 times on the base sensors on each side of the OAVs, although most shocks did not correlate in time between sensors.  The few sensor-correlated shocks seem to correspond to periods of loading or unloading.  Periodic vertical acceleration measurements are shown in Figure~\ref{fig:accel}; during the trip, typical accelerations were between 0.2~g and 0.4~g in excess of gravity.  The changes in acceleration patterns give a picture of the status of the vessel, whether it was moving on the road, at rest, moving on a ship, or being loaded and unloaded.

\begin{figure}[htpb]
\centering
\subfigure[Temperature measurements during OAV transport. The figure is split into five periods: by truck in USA (red), at port in Long Beach (yellow), at sea (green), at port in Yantian (gray), and en route to and in the SAB (purple).] {
  \includegraphics[trim=0cm 0cm 15cm 12.5cm, clip=true, width=0.46\textwidth]{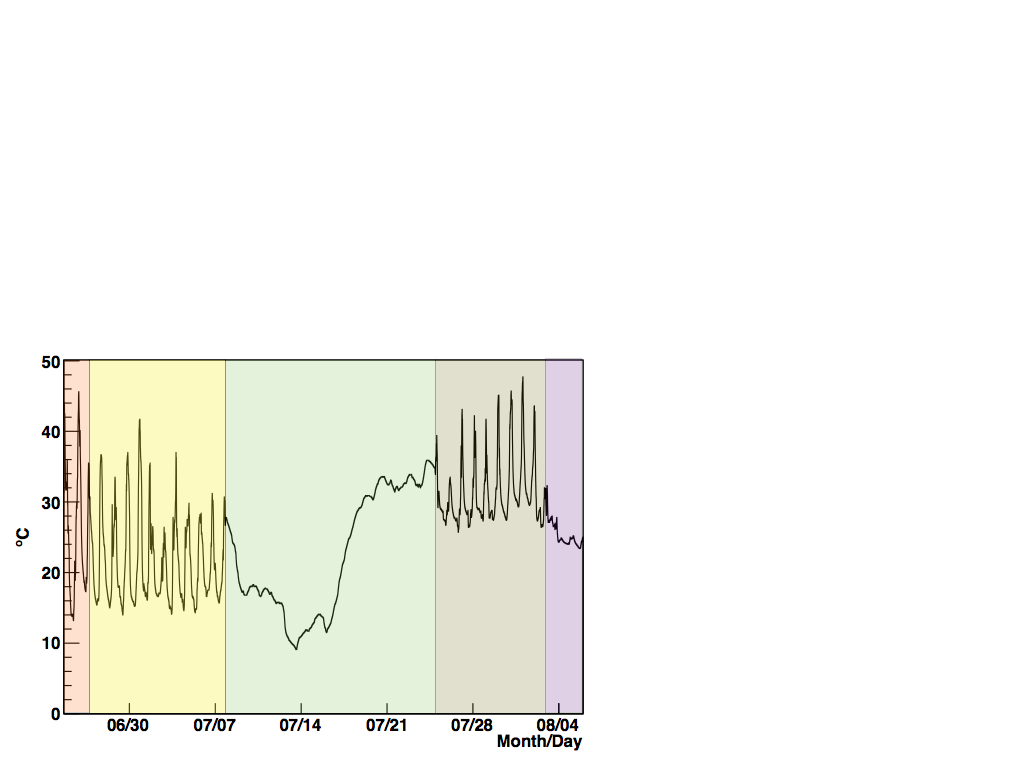}
  \label{subfig:temp}
}
\hspace{0.1cm}
\subfigure[Pressure measurements during OAV trucking transport in the US.  Pressure changes are due to altitude variations along the trucking route shown in Figure~15.  The gray-shaded periods correspond to times when the OAVs were at rest.] {
  \includegraphics[trim=8cm 7cm 10cm 7.5cm, clip=true, width=0.46\textwidth]{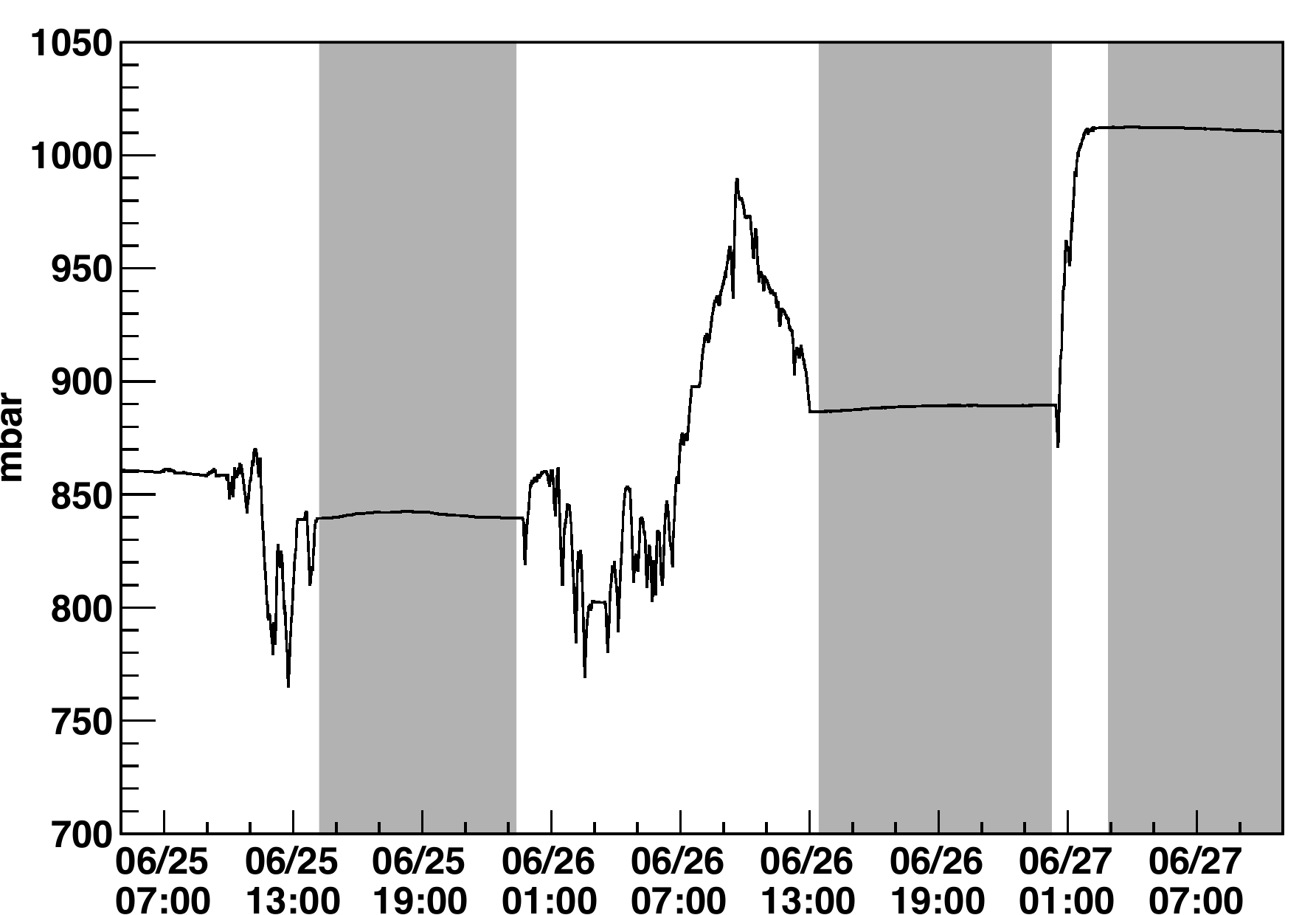}
  \label{subfig:pressure}
}
\caption{Temperature and pressure of the OAV environment during transport. Note that the changes of temperature and pressure with time convey the vessel's current general location.}
\label{fig:Env}
\end{figure}

The temperatures over time recorded by one sensor on the OAV shipping frame base is shown in Figure~\ref{subfig:temp}.  Periods of rapid temperature fluctuation correspond to time spent outside on the road or in port, while the smoothly varying period corresponds to time spent on the ocean.  The temperature sensor recorded variations between 8~$^{\circ}$C and 48~$^{\circ}$C, with a maximum of 29~$^{\circ}$C change over the course of 12 hours.  Another temperature sensor located within the OAV packaging on its lid experienced temperature variations of only 20~$^{\circ}$C.

This 20~$^{\circ}$C temperature fluctuation, if also experienced by the OAV, corresponds to a 4~mm variation in radius of the OAV from thermal expansion and contraction.  It was not clear from the available temperature data how much temperature fluctuation, and thus acrylic expansion and contraction, varied over the surface of the OAV.  Excessive or uneven expansion and contraction of the OAV is a concern as it is not known if such behavior will affect the mechanical strength of the OAV appreciably.  Temperature tests have been conducted that ensure acrylic and bond optical and mechanical stability in the presence of these environmental fluctuations.  As an added precaution, future OAVs were shipped with extra insulation to flatten out these temperature fluctuations.

The recorded pressure history during the driving portion of the trip is shown in Figure~\ref{subfig:pressure}; the rise in pressure at the end of the trip is the result of the transport truck coming down to sea level.  Differences in pressure inside and outside an OAV were never greater than 1~mbar.  The light sensor data did not show any unexpected periods of brightness.  In addition, a light-sensitive acrylic monitor sample mounted by the OAV1 lid sensors did not show any signs of optical degradation, as would be expected if it had experienced any moderate exposure to sunlight.

\section{Assembly}

\subsection{Cleanliness and Cleaning}

When they reach Daya Bay, acrylic vessels must be cleaned to ensure that the amount of radioactive contaminating material inside the detector's target region is minimized.  In addition, cleaning improves the optical quality of the IAV and OAV surfaces and removes contaminants that are incompatible with the AD liquids.  The quantitative cleanliness goal is to ensure that the radioactivity rate from contaminants in the IAV and OAV is less than 10\% of the expected radioactivity from the GdLS liquid.  This corresponds to a contamination radioactivity of 1~Hz.  Table~\ref{tab:Dirty} presents rough estimates for acceptable amount of possible contaminants based on their activity.

\begin{table}[htpb]
\centering
\begin{tabular}{|c|c|c|c|}
\hline 
Contaminant & Source & Isotope & Acceptable Amount \\ \hline \hline
Al$_2$O$_3$ Powder & Spot-Scrubbing & $^{40}$K &  36 kg \\ \hline
Fingerprints & Human Contact & $^{40}$K & > 1000 handprints \\ \hline
Human Sweat & Human Contact & $^{40}$K &  < 1 cm$^3$ \\ \hline
Bamboo or Wood & Used for Scaffolding & $^{40}$K &  < 3 cm$^3$ \\ \hline
Dirt & Surrounding Environment & $^{40}$K,$^{232}$Th,$^{238}$U &  2 g \\ \hline
Paint Chips & Semi-Cleanroom Crane & $^{40}$K,$^{232}$Th,$^{238}$U & < 50 g \\ \hline
\end{tabular}
\caption{Possible radioactive contaminants in the OAVs and IAVs, listed with their maximum acceptable post-cleaning presence.}
\label{tab:Dirty}
\end{table}

The OAVs and IAVs were cleaned at the times listed in Table~\ref{tab:timeline}.  Cleaning of the IAV insides was done over the course of two days at Nakano.  Cleaning of the IAV outsides was done in the course of a day per IAV at the Daya Bay SAB.  OAV cleaning took 4-5 days per vessel at the SAB.

\begin{figure}[htpb]
\centering
\subfigure[Scrubbing the OAV inside with 1\% Alconox solution.] {
  \includegraphics[width=0.5\textwidth]{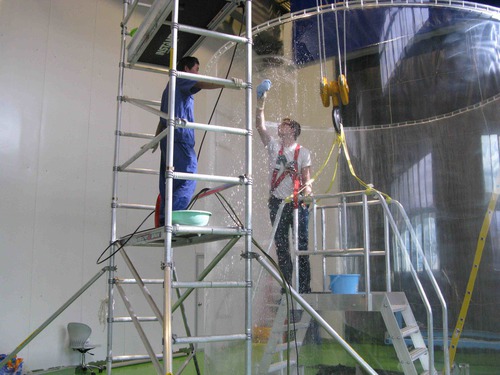}
  \label{subfig:Cleaning1}
}
\hspace{0.1cm}
\subfigure[Testing rinse water conductivity with a conductivity meter.  All conductivity meters used were accurate to 0.1 $\mu$S.] {
  \includegraphics[width=0.43\textwidth]{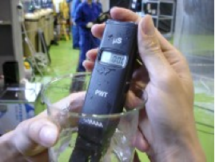}
  \label{subfig:Cleaning2}
}
\caption{Photographs taken during IAV and OAV cleaning in the SAB at Daya Bay.}
\label{fig:Cleaning}
\end{figure}

Cleaning of the OAVs is done in the SAB semi-cleanroom, a climate-controlled, low-particulate environment.  The top of the vessel is removed, and the top flange and lid are then spot-scrubbed with Al$_2$O$_3$ powder, water, and a microwipe cloth to remove salt water residue, rust, and plastic wrap residue.  The entire lid is washed on top and bottom with a 1\% Alconox solution and fresh microwipes and rinsed with a pressure washer using 10~M$\Omega$ water.  The conductivity of the rinse water is measured with a conductivity meter.  When the rinse water conductivity matches that of the water directly from the pressure washer, rinsing is considered complete.  The lid is then dried with microwipe towels and covered with a clean tarp.  This same process is applied to the inside and outside of the OAV.  After cleaning the vessel inside and walls, the lid is then reconnected and the vessel is lifted onto blocks so that the bottom can be spot-scrubbed and cleaned.  A few pictures of the cleaning process can be seen in Figure~\ref{fig:Cleaning}.

The same general procedure is applied during the cleaning of the outside of the IAV and the acrylic overflow tank parts in the SAB semi-cleanroom.  The inside of the IAV is also cleaned in the same manner, but in a class 10,000 clean tent at the Nakano factory before the vessel is completely bonded together.  AV tube connection hardware parts are cleaned using an ultrasonic cleaner first in a 1\% Alconox solution, and then in water.

\subsection{Leak-Checking}

Vessels are leak-checked to ensure that the target volume remains well-known to better than 0.1\% after five years of detector operation.  In addition, the leaking of non-scintillating liquids like water or mineral oil into the scintillating regions will reduce the detector's light yield.

The leak requirements for the inner volumes are listed in Table~\ref{tab:Leaks}; if these requirements are met, the uncertainty in target mass resulting from leakage will be $<$0.013\%.  Leak rates of argon in cc/sec at a given pressure differential are calculated from the 5-year acceptable leakages to serve as specifications for the leak-checking process.  While the entire AD is surrounded by water, no single direct water-to-LS junction exists.  Because of this, a water-to-LS leak rate specification is not applied.

\begin{table}[htpb]
\centering
\begin{tabular}{|c|c|c|c|}
\hline 
Leak Type & $\Delta$P (cm H$_2$O) & 5~yr Leakage (l) & Argon Leak Rate (cc/sec) \\ \hline \hline
LS to or from Gd-LS & 15 & 3 & 3.8$\times10^{-2}$ \\ \hline
MO to LS & 15 & 5 & 6.4$\times10^{-2}$ \\ \hline
\end{tabular}
\caption{Leak rate specifications for the central liquid volumes, given for argon and for the AD liquids.  As the leak rate is dependent on the pressure differential between zones, a conservative estimate of the zone pressure differential is also given.}
\label{tab:Leaks}
\end{table}

Each AV pair has numerous seals in four main locations: on the OAV lid, on the IAV calibration/overflow tube stacks, on the OAV calibration/overflow tube stacks, and near the stainless steel lid connections.  Some leak-testing procedures are done at Reynolds, and some are done on-site at the Daya Bay SAB using varied procedures that depend on the type and location of the seal.

Double o-ring seals are used facilitate high-precision leak checking of the critical OAV flange connection.  This seal can be checked by pressurizing the region between the o-ring seals with argon using a special gas input port, and then monitoring any change in pressure over time.  The maximum leak rate on the main seal of OAV1 and OAV2 was 4.4$\times10^{-2}$~cc/sec at a pressure of around 5~psi.  This is well below 0.1~cc/sec,  the maximum acceptable argon leak rate from MO to LS given in Table~\ref{tab:Leaks}, corrected to a pressure of 5~psi.

\begin{figure}[htb]
\centering
\includegraphics[width=3.5 in]{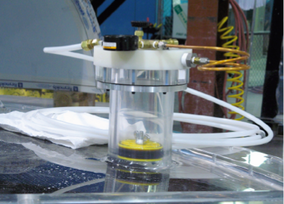}
\caption{Photograph of the double o-ring seal leak-checking setup for the OAV ports.  The yellow seal at the bottom of the port is clearly visible, as is the gas input.}
\label{fig:leakreynolds}
\end{figure}

The OAV port single o-ring seals were tested at Reynolds by placing a tight seal at the bottom of the port, as in Figure~\ref{fig:leakreynolds}, and then pressurizing the entire sealed port to 5~psi and monitoring any change in pressure with a pressure gauge.  The maximum change in pressure for any OAV port was 3.2$\times10^{-2}$~cc/sec, significantly below the 0.1~cc/sec leak-rate limit.

\begin{figure}[htpb]
\centering
\subfigure[Diagram of the gas circuit for AV leak checking.  Gas is fed from bottles through a gas control rack and the into AV.  Output air is pushed out of the AV and through an exhaust line to the exterior of the SAB.  AV gas pressure is controlled through the use of pressure relief bubblers.] {
  \includegraphics[trim=0.1cm 9.5cm 19cm 2.5cm, clip=true, width=0.50\textwidth]{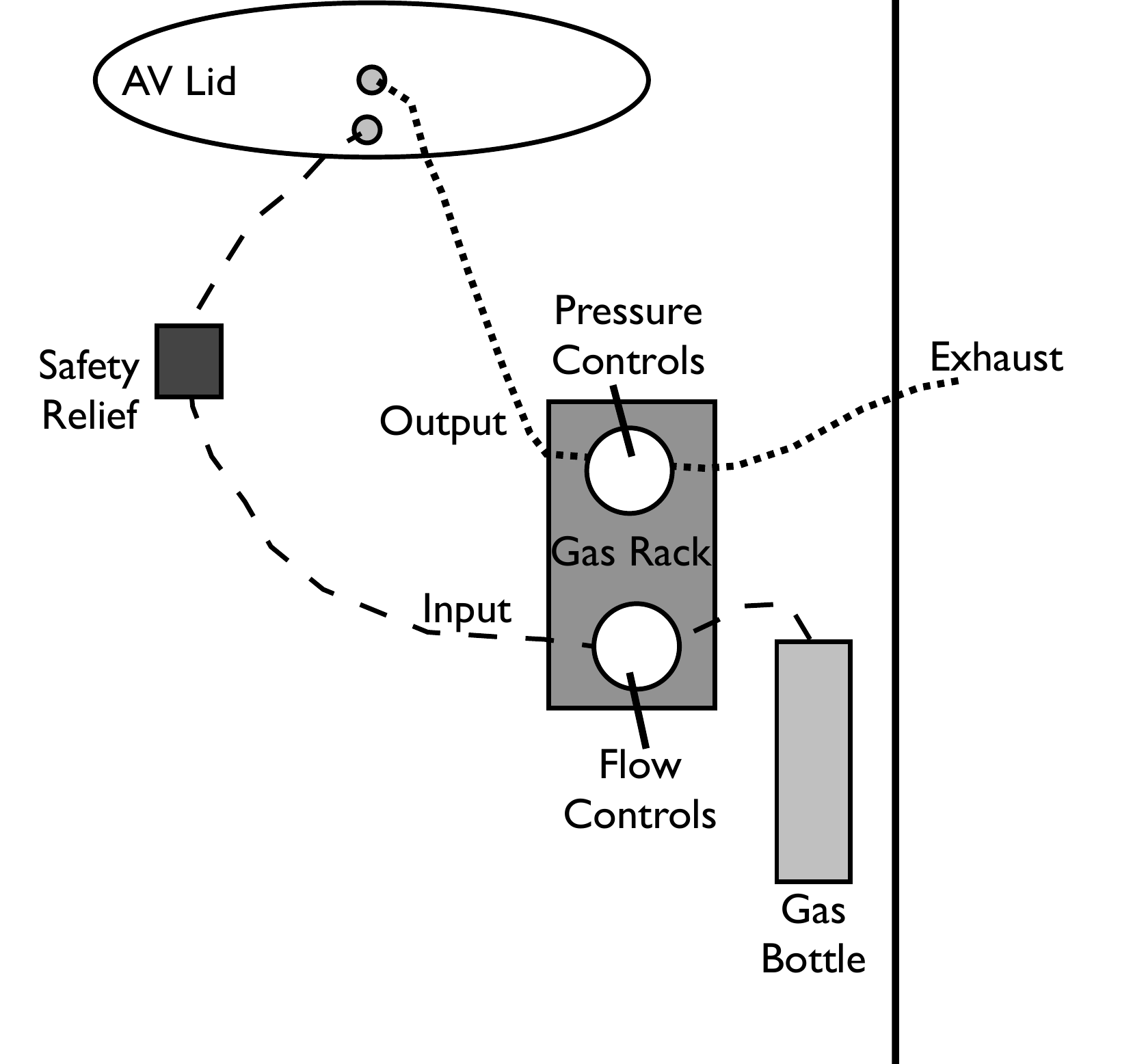}
  \label{subfig:LeakDiagram}
}
\hspace{0.5cm}
\subfigure[Photograph of the IAV tube stack leak-checking setup.  An argon sniffer is run up and down the tube stack, which is pressurized with argon gas.] {
  \includegraphics[width=0.40\textwidth]{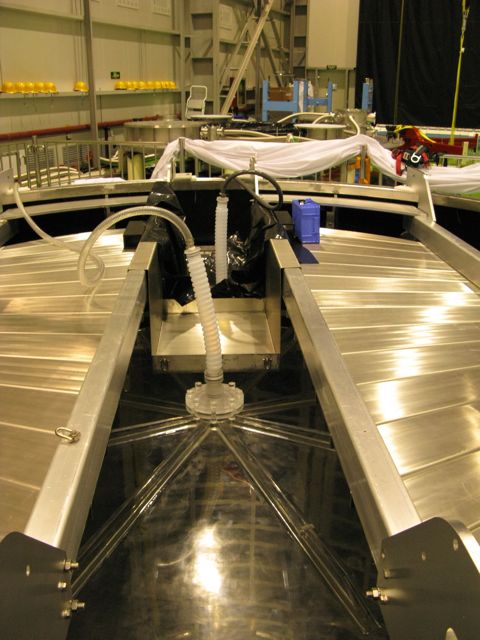}
  \label{subfig:LeakSAB}
}
\caption{An overview of the gas circuit and appearance of the leak-checking system in the Daya Bay SAB cleanroom.}
\label{fig:LeakCheck}
\end{figure}

For single and double o-ring seals that cannot be easily accessed or isolated, a different leak-checking procedure is used.  The calibration tube stacks are assembled and attached to the AV, which is then pressurized to 10-15~cm water column with argon gas.  Gas input routing and pressure control are centralized in a gas rack located in the SAB cleanroom.  Gas output from the AD is run through an exhaust system to the SAB exterior.  To prevent AV overpressuring from line kinks and other gas system malfunctions, safety relief valves are also placed at a number of locations in the gas system.  A diagram of the gas system circuit can be seen in Figure~\ref{subfig:LeakDiagram}.  When the AVs are sufficiently pressurized, an argon sniffer can be used to check for leaks along the tube stack.   The appearance of the IAV lid area during IAV tube stack leak checking can be seen in Figure~\ref{subfig:LeakSAB}.  The sensitivity of this method is a minimum leak rate of 10$^{-3}$~cc/sec and is limited by the stability of the gas sniffer.

The maximum measured leak rate measured using this method was 3$\times10^{-3}$~cc/sec at one location on on OAV1, well within the maximum specified Gd-LS-to-LS leak rate listed in Table~\ref{tab:Leaks}.  The leak rates in all other areas were below the sensitivity of the gas sniffer.  The same method is used for measuring the leak rates in the OAV calibration tube stacks with similar sensitivity and results.

A final test of the aggregate leak-tightness of the AVs was made after completing installation of the IAV and OAV tube stacks and SSV lid.   Using the same gas circuit from the previous test, the Gd-LS and MO regions were pressurized with argon at 10~cm water pressure, while the LS region was filled with freon at 1~cm water pressure, to ensure minimal freon leakage during filling.  Freon and argon concentrations were verified by measuring the lowered oxygen content of the exhaust air from the regions with an oxygen monitor.  Once the LS region was composed of $>$75\% freon, the pressure differentials were reversed, leaving the freon-filled LS volume pressurized with respect to the Gd-LS and MO volumes; this maximized any possible freon leakage. After 20 hours, the freon concentration of the AD1 Gd-LS and MO region exhausts were measured at 66~ppm and 52~ppm with a freon sensor; similar values were found for AD2.  These values are below 159~ppm and 133~ppm, the maximum values of acceptable freon concentration in the Gd-LS and MO exhaust based on the AD leakage requirements.

As was mentioned before, the acrylic overflow tank components are contained in a nitrogen gas environment.  The leak-tight stainless steel housing that separates this gas volume from the water volume surrounding the AD and overflow region will not be discussed in this paper.  Any possible leakage or splashing between overflow tank volumes is minimized by the acrylic covers on each volume.

\subsection{Installation}

\begin{figure}[htpb]
\centering
\subfigure[Drawing of a lifted OAV.] {
  \includegraphics[width=0.45\textwidth]{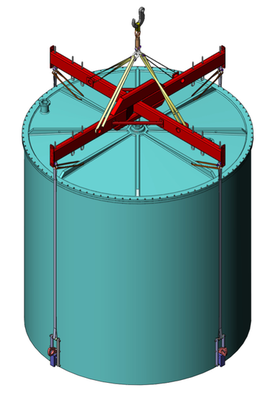}
  \label{subfig:OAVLift}
}
\hspace{0.5cm}
\subfigure[Photograph of the OAV just before lifting in the SAB.] {
  \includegraphics[width=0.45\textwidth]{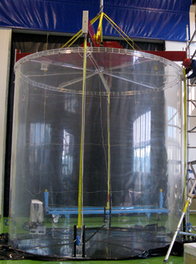}
  \label{subfig:}
}
\caption{A drawing and photograph of the OAV rigged for lifting.  Visible are the lifting fixture and the rigging connected to the OAV via the OAV lifting hooks.}
\label{fig:OAVLift}
\end{figure}

After the OAV is cleaned in the semi-cleanroom, the OAV lid is temporarily reattached and the OAV is moved into the cleanroom.  Using the 40~ton cleanroom crane and proper rigging, pictured in Figure~\ref{fig:OAVLift}, the OAV is lifted and lowered into a cleaned SSV located in the AD assembly pit.  The OAV is set down on the bottom reflector, a 4.5~m diameter circular sheet of specularly reflecting material sandwiched between two 10~mm thick acrylic sheets.  The reflector rests on the SSV bottom ribs.  Surveys are then done on the OAV lid using a Leica System 1200 Total Station to determine if the OAV is level and if the OAV ports line up with the previously surveyed locations of the SSV ports~\cite{Leica}.  The vessel is shimmed, re-situated, and re-surveyed until the SSV and OAV ports are acceptably aligned and levelled.  Survey details and results will be further discussed in Section 6.3.6.  When surveys are complete, stainless steel hold-down mechanisms are installed over the OAV lifting hooks to hold the OAV in place during all future AD activities.

\begin{figure}[htb]
\centering
\includegraphics[width=5 in]{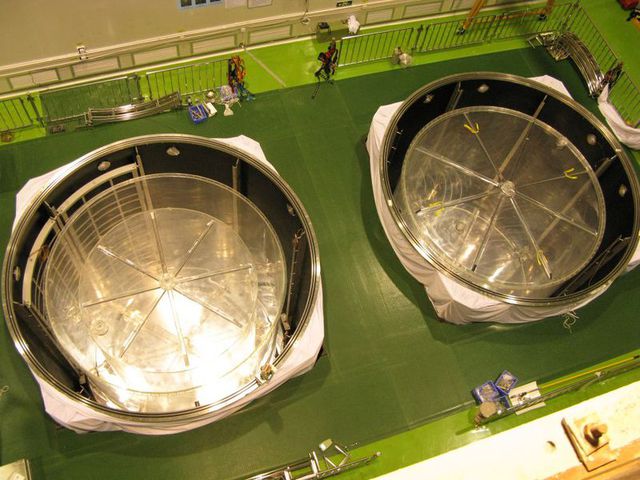}
\caption{A photograph of the assembly of a pair of ADs in the cleanroom.  Assembly for the two vessels is at different stages in the photo: AD1, on the left, is shown with a nested IAV and OAV, while AD2, on the right, is shown with an OAV with lid attached. }
\label{fig:Nested}
\end{figure}

Once the OAV is in place, its lid is removed and the OAV flange is surveyed.  Shims are then installed along with the IAV support pucks, to ensure that the IAV is level when installed.  A cleaned IAV is lifted using the same lifting frame and slightly different rigging, and lowered into the OAV until it is resting on the pucks.  The IAV ports and top are then surveyed to check if they are concentric with the other OAV and SSV ports, and repositioned and shimmed if necessary.  This data is also included in the Section 6.3.6.  Once proper alignment has been achieved, the hold-down mechanisms are engaged to secure the IAV to the OAV.  A final survey of the IAV top is recorded to provide flatness and concentricity data for geometric characterization.  A photo of the nested acrylic vessels can be seen in Figure~\ref{fig:Nested}.  Assembled ladders of PMTs are installed after the securing of the vessels.

At this point, the IAV connection tubes are installed, leak-checked, and removed so that the OAV lid can be installed and its double o-ring seal leak-checked.  The top AD reflector, which rests on top of the OAV lid, can then be installed.  Next, OAV connection tubes are installed onto the OAV lid and are leak-checked.  If all seals pass their leak-checks the SSV lid can then be installed.  Once the SSV lid is secure, off-center OAV connection tubes are secured to the SSV, and IAV connection tubes are fed down through the OAV tubes and connected to the IAV ports.  With the the lid and connection tubes properly installed, the final three-zone leak check can be performed. The overflow tank assembly is then assembled and connected to the central OAV and IAV tubes and SSV lid, completing the AV and overflow tank installation.  A picture of overflow tank installation can be seen in Figure~\ref{subfig:OverflowInstall}.

\section{Acrylic Vessel Characterization and QA  Tests}

\begin{figure}[htb]
\centering
\includegraphics[trim=0.5cm 8cm 0.2cm 3.5cm, clip=true, width=5.5 in]{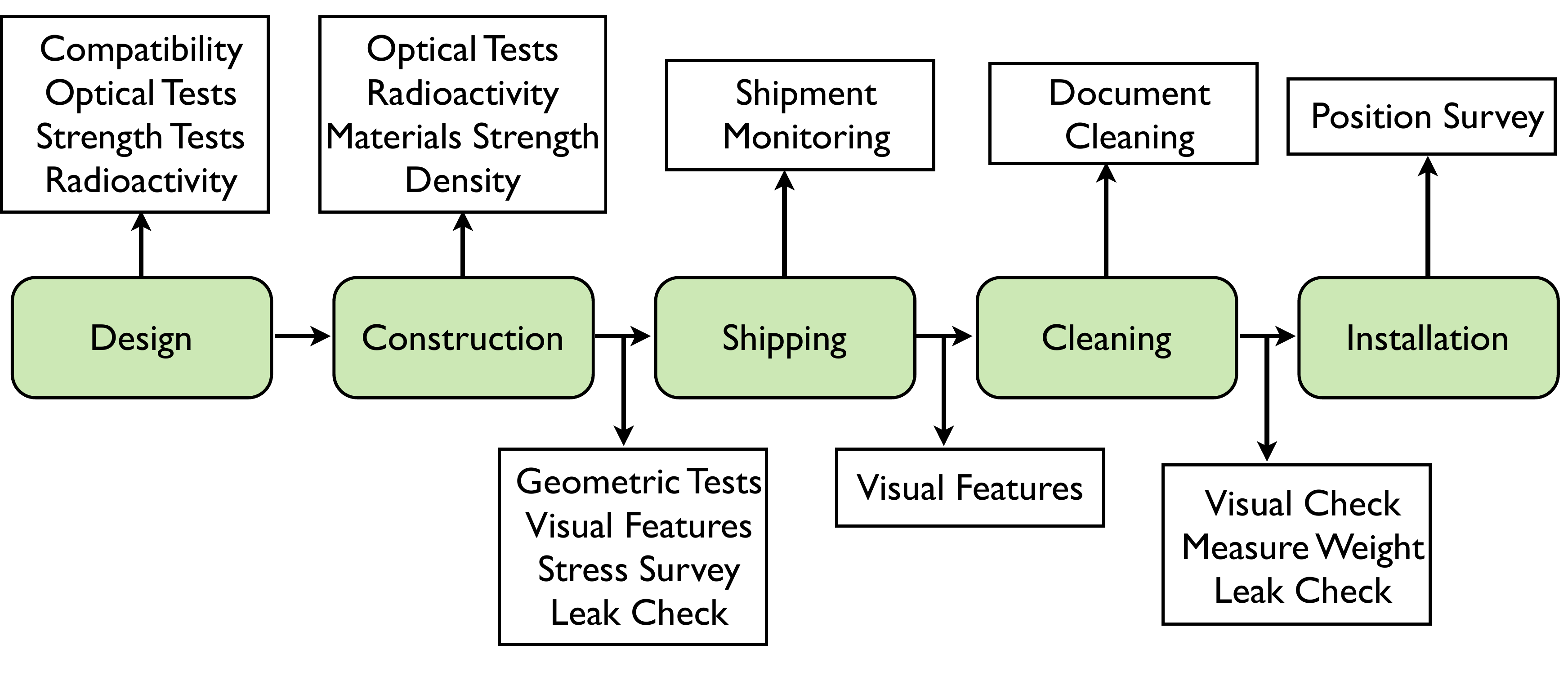}
\caption{A flow chart depicting what type of characterization data is collected during the various stages of the acrylic vessels' life cycles.  The rounded rectangles depict the life cycle stages, while rectangles describe collected QA and characterization data.}
\label{fig:lifedatachart}
\end{figure}

The acrylic vessels must undergo quality assurance testing to ensure that the vessels meet all engineering and physics requirements previously mentioned in this paper.  The QA program described here measures all possible geometric, optical, and material properties relevant to engineering and physics considerations, as well as documenting all stages of the AV life cycles.  Table~\ref{tab:testtypes} summarizes the measurements that have been made, and Figure~\ref{fig:lifedatachart} describes when during the AV life cycle each measurement was made.  In this section, each AV characterization measurement will have its methods described and its results summarized.

\begin{table}[htpb]
\centering
\begin{tabular}{|c|c|}
\hline 
Category & Vessel Property \\ \hline \hline
\multirow{6}{*}{Optical} & Transmittance \\ \cline{2-2}
 & Index of Refraction  \\ \cline{2-2}
 & Attenuation Length  \\ \cline{2-2}
 & Visual Appearance \\ \cline{2-2}
 & Surface Quality \\ \cline{2-2}
 & UV Degradation \\ \hline
\multirow{4}{*}{Geometric} & Inner/Outer Diameter  \\ \cline{2-2}
 & Thickness  \\ \cline{2-2}
 & Plumb Bob (Perpendicular Sides) \\  \cline{2-2}
 & O-Ring Groove Depth  \\ \cline{2-2}
 & Positioning: Survey Data  \\ \hline
\multirow{3}{*}{Material Properties} & Radioactivity \\ \cline{2-2}
 & Chemical Compatibility \\ \cline{2-2}
 & Material Strength \\ \cline{2-2}
 & Density  \\ \hline
\multirow{2}{*}{Quantity} & Weight \\ \cline{2-2}
 & Volume \\ \hline
\end{tabular}
\caption{Overview of vessel properties measured for quality assurance and characterization.}
\label{tab:testtypes}
\end{table}

Tests on the vessels were performed at Nakano, Reynolds, or at the SAB.   Many tests on the raw material samples were conducted in labs at University of Wisconsin - Madison or National Taiwan University.  Radioactivity measurements were done by Berkeley National Laboratory and California Institute of Technology, and the bulk of chemical compatibility information was provided by studies at Brookhaven National Laboratory.

\subsection{Design Requirements and Materials Selection}

Initial quality assurance and R\&D measurements were done during the design stage to determine the correct material with which to build the acrylic vessels.  Of particular concern were the optical, radioactive, and mechanical properties of candidate materials, and their compatibility with prospective target and gamma catcher liquids.

\subsubsection{Optical Properties}

The transparency requirements of the acrylics used in the IAV and OAV are dictated by the emission and transmittance spectrum of the AD liquids, as well as by the quantum efficiency spectrum of the Daya Bay PMTs \cite{Bryce-2009, Hamamatsu}.  From these inputs, it was determined that the IAV and OAV acrylic needs to be highly transparent to photons with wavelengths from 360~nm to 500~nm.  The specifications on transmittance for a 10-15~mm sample in air for IAV and OAV acrylics for Daya Bay were 84\% at 360~nm, 88\% at 380~nm, 90.5\% at 400~nm, and 91.5\% at 500~nm.  These values mirror those used in the SNO experiment.  Acrylic sheets with transmittances of a few percent below specifications for lower wavelengths were also deemed acceptable as a majority of the scintillation light propagated in the Daya Bay liquids is above 400~nm.

During the design phase, a variety of samples were tested using an SI Photonics Model 440 UV-Vis Spectrometer \cite{SIPhotonics} to identify acrylic types that met these transmittance specifications.  UVT acrylic meets these requirements, while standard UVA acrylics only begin transmitting light appreciably at 400~nm.  Eventually, thin UVT acrylic sheets from Polycast were selected for the OAV walls, cast blocks of UVT acrylic from Reynolds were selected for the OAV lid and bottom, and PoSiang UVT acrylic was selected for the IAV.  UVA acrylic was used for the overflow tanks and connection hardware, as these components are either small or far removed from the detector target.  Transmittance data from the UVT acrylics are presented in Section 6.2.3.

\subsubsection{Radioactivity Testing}

Acrylic is an organic compound and does not contain many of the long-lived radioactive isotopes that are present in other optical window materials, such as glass. To ensure that radioactive backgrounds from the selected production acrylic types were acceptably low, samples of each type of acrylic were counted for radioactivity using a high-efficiency HPGe detector situated inside a low-background external radioactivity shield.  Measurements were made with similar setups at either a concrete-shielded surface facility at Berkeley National Laboratory or at an underground laboratory at the Oroville Dam in Oroville, California~\cite{Oroville}.  Radioassay results are listed in Table~\ref{tab:radioactivity}.  Note that the value for all measurements is only an upper bound, which is limited by the sensitivity of the equipment, the size of the sample, and the duration of the assay.

In order to keep singles rates and correlated backgrounds at an acceptable level, it is desired that the intrinsic bulk radioactivity contribution from each AV component be lower than the radioactivity requirement for the 20~kg of Gadolinium salt used to manufacture the Gd-LS.  Table~\ref{tab:radioactivity} includes the measured radioactive isotope concentration upper limits for each acrylic type, as well as the isotope concentrations that will achieve the radioactivity requirement.  For Polycast acrylic and for $^{40}$K concentrations, these requirements have been met by the measurements.  More precise measurements have been planned to lower the limits for Reynolds and PoSiang acrylics, so that low backgrounds in these acrylics can be confirmed.

\begin{table}[htp]
\centering
\begin{tabular}{|c|c|c|c|c|c|c|}
\hline 
\multirow{2}{*}{Acrylic Type}  & \multicolumn{2}{|c|}{$^{238}$U (ppb)} & \multicolumn{2}{|c|}{$^{232}$Th (ppb)} & \multicolumn{2}{|c|}{$^{40}$K (ppm)} \\ \cline{2-7}
& Measured & Goal & Measured & Goal & Measured & Goal \\ \hline \hline
Reynolds  & <3 & 0.1 & <3 & 0.3 & <1 & 13 \\ \hline
Polycast  & <0.07 & 0.09 & <0.2 & 0.2 & <0.1 & 9 \\ \hline
Posiang  & <0.2 & 0.1 & <0.4 & 0.2 & <0.3 & 11\\ \hline
\end{tabular}
\caption{Measured and required limits on concentration of radioactive isotopes in the production AV acrylics. }
\label{tab:radioactivity}
\end{table}

\subsubsection{Materials Compatibility}

\begin{figure}[htpb]
\centering
\subfigure[Setup of mechanical compatibility tests of various candidate target liquids with acrylic.] {
  \includegraphics[width=0.62\textwidth]{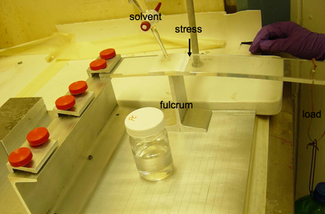}
  \label{subfig:StressCompat}
}
\hspace{0.1cm}
\subfigure[Photograph of an acrylic stress testing setup.  The acrylic is the long clear bar connecting the top metal hitch to the bottom weight] {
  \includegraphics[width=0.31\textwidth]{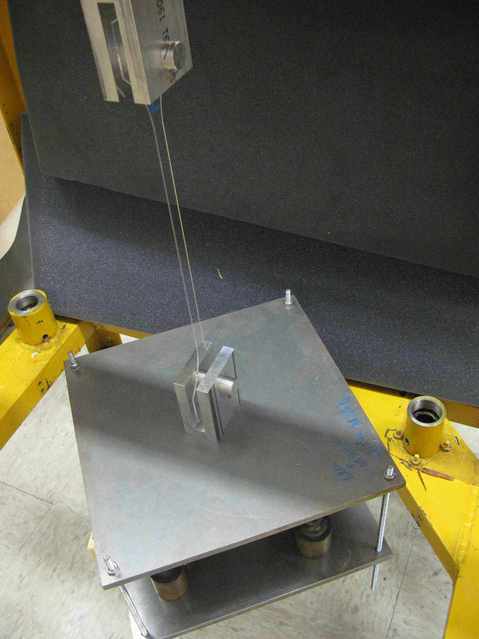}
  \label{subfig:StressSetup}
}
\caption{Photographs of two different stress testing setups, one for testing material compatibility, and one of testing maximum stress limits.}
\label{fig:Stress}
\end{figure}

It is important to test optical and mechanical compatibility between acrylic and candidate detector liquids, as many solvents and some liquid scintillators are known to cause crazing in stressed regions in acrylic.  Mechanical compatibility testing was carried out in a manner described in the Handbook of Acrylics, the standard reference text on acrylics \cite{Stachiw}; figures of the test setup can be seen in Figure~\ref{subfig:StressCompat}.  Strips of acrylic of well-known dimensions were held in position at one end and stressed on the other with a known weight.  A fulcrum was placed underneath the middle of the acrylic strip; above the acrylic in the area of the fulcrum a filter paper was placed and saturated with the test liquid.  The stressed acrylic was then observed over a period of time to observe any changes resulting from the liquid-acrylic contact.  Pseudocumene, a liquid scintillator solvent used in other experiments such as KamLAND, caused the stressed acrylic piece to break within one hour of loading the acrylic.  Linear alkyl benzene (LAB), a commonly used solvent for detergents, did not have any appreciable effect on the acrylic after 30 hours of stressing, making it a good candidate use at Daya Bay.

Optical compatibility testing was done by placing acrylic samples into samples of candidate liquids for extended periods of time at elevated temperatures, to accelerate any possible leaching or other chemical interactions between the acrylic and liquid.  The degradation rate with time of the LAB transmittance when exposed to acrylic was negligibly small after 21-28 days of testing at 40$^{\circ}$C. as LAB was the most optically and mechanically compatible liquid with acrylic, it was selected as the liquid scintillator for the production target and gamma catcher liquids.

\subsubsection{Mechanical Strength}

To ensure that the AVs are structurally sound over the life of the experiment, the vessels, like most other large acrylic structures, are designed for a maximum stress of 5 MPa and a lifetime of 10 years.  A setup was constructed to independently test these limits by stressing acrylic samples at known amounts and observe the effects over time.  A photograph of the setup is shown in Figure~\ref{subfig:StressSetup}.  Long, thin acrylic pieces of well-known cross-section are hung from a frame from one end while known weights are hung from the bottom of the acrylic, which creates uniform, well-known stresses in the main length of the acrylic.

The first trial consisted of an acrylic strip stressed to 24~MPa.  Significant crazing was viewed in the piece within 24 hours, and the piece broke after 48 hours.  The second strip was stressed to 15~MPa, and moderate crazing was visible after 8 days of hanging.  This piece was then removed and replaced with another strip that was stressed to 10 MPa.  Minor crazing appeared after about 6 months and has persisted but not worsened appreciably after 2 full years of hanging.  From these trials, the figure of 5~MPa over 10 years seems acceptable; vessels were designed to be supported such that stresses were well below the 5~MPa limit for all parts of the vessel structure.  In addition, these tests show that short-term (on the order of a few days) stresses around 10-15~MPa are acceptable; this figure provides a guide to acceptable maximum stresses allowed during short-term processes such as shipping, lifting, and filling.

\subsection{Optical and Mechanical Characterization}

\subsubsection{Visual Inspection}

\begin{figure}[htb]
\centering
\includegraphics[width=3 in]{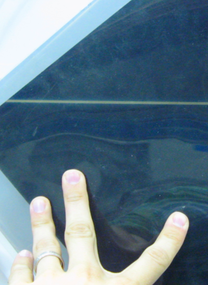}
\caption{Photograph of a horizontal bond-line, viewed through perpendicular polarized filters.  Microcrazing is present above this bond line but its visible appearance is masked by the polarized filters.}
\label{fig:Bond}
\end{figure}

By visually inspecting the acrylic vessels, features and defects that alter the acrylic's clear, completely transparent appearance can be identified.  Careful visual inspection reveals bond lines where acrylic sheets are joined together with bonding syrup.  Identification of bonds is made easier by placing polarized filters on either side of the vessel that are aligned perpendicularly.  In this configuration, birefringence effects cause bonds to appear lighter in color than the surrounding acrylic.  The joining of two acrylic sheets is pictured in Figure~\ref{fig:Bond}.  By comparing the visible bond line pattern to the designed vessel plans, it was confirmed that the vessels were constructed as designed.

\begin{figure}[htb]
\centering
\includegraphics[width=4.5 in]{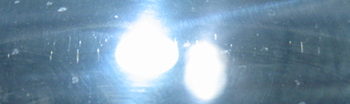}
\caption{Microcrazing visible above the equator bond of OAV2.  Backlighting and indirect-angle viewing of the crazed area makes it easier to view.}
\label{fig:crazing}
\end{figure}

Visual features are largely found along bond lines.  Isolated groups of small bubbles caused by the exothermic curing reaction are visible in a few spots, especially along the middle horizontal bond joining the two vessel barrel cylinders.  Short vertical crazing marks, mostly less than 1/4'' long and less than 1/16'' deep were also visible at places on or just above the same bond; an example is shown in Figure~\ref{fig:crazing}.  The cause of this minor crazing is likely excess pressure placed on the inside of the wall cylinders while bonding them together: in order to hold the bonding syrup between the cylinders, dams needed to be pushed firmly against the sides of the walls.  These slightly crazed areas are not a concern, as the stress that caused them is no longer being applied.  In addition, they occur in areas of low expected stress under normal conditions.  Areas of more significant crazing were sanded out and filled with bonding syrup.  Minor chips and bonding and machining features were also visible but not worrisome.

\subsubsection{Surface Properties}

\begin{figure}[thpb]
\centering
\includegraphics[width=3 in]{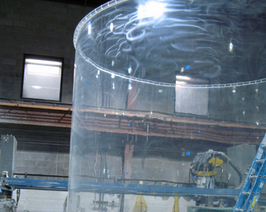}
\caption{Surface appearance of OAV.  Halos of diffusely reflected light are visible at the top of the OAV.  Despite this, the overall surface quality of the vessel is good, giving the vessel a largely transparent appearance.}
\label{fig:LightHalo}
\end{figure}

\begin{figure}[htpb]
\centering
\subfigure[Acrylic sample polished to 400-grit, in air.] {
  \includegraphics[width=0.35\textwidth]{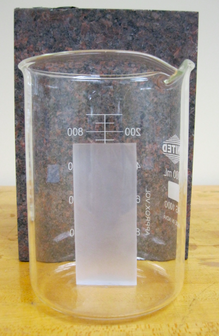}
  \label{subfig:airimmersed}
}
\hspace{1cm}
\subfigure[Acrylic sample polished to 400-grit, immersed in mineral oil.] {
  \includegraphics[width=0.35\textwidth]{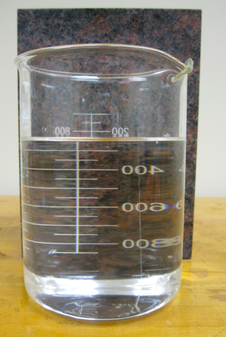}
  \label{subfig:MOimmersed}
}
\caption{A comparison of acrylic surface appearance when surrounded by two mediums of different refractive index, air (n$\sim$1) and mineral oil (n$\sim$1.36).  Acrylic has n$\sim$1.50}
\label{fig:immersed}
\end{figure}

The surface of OAVs and IAVs were polished to 1~to~3~micron grit by Reynolds and Nakano.  This gives the vessels a largely transparent finish with small halos of diffusely reflected light visible from certain perspectives, as seen in Figure~\ref{fig:LightHalo}.  The IAVs appear to have a slightly more diffusive surface than the OAVs.  However, these surface effects will largely disappear when the vessels are immersed in liquids.  Figure~\ref{fig:immersed} illustrates the disappearance of surface features and roughness when acrylic samples are immersed in liquids of similar index of refraction.  Exact index of refraction values will be discussed in Section 6.2.4.

The only minor surface defect present in the acrylic vessels was on OAV1.  The Genmac plastic wrap used to protect the outer surface of the vessel during shipping was left on the OAV1 bottom after unpackaging.  Over four months of storage, the plastic wrap degraded, leaving a sticky film in spots on the bottom of the vessel.  In lieu of using nonpolar compounds whose compatibility with the acrylic was unknown, brass brushes were used to abrade the surface of the OAV and remove the sticky residue.  The surfaces in these abraded areas is more diffusive than other areas, while still mostly transparent.  We qualitatively estimate that 10-20\% of the bottom surface of OAV1 is abraded.  These spots disappeared as OAV1 was filled with liquid.

\subsubsection{Transmission Spectrum}

\begin{figure}[htb]
\centering
\includegraphics[width=4 in]{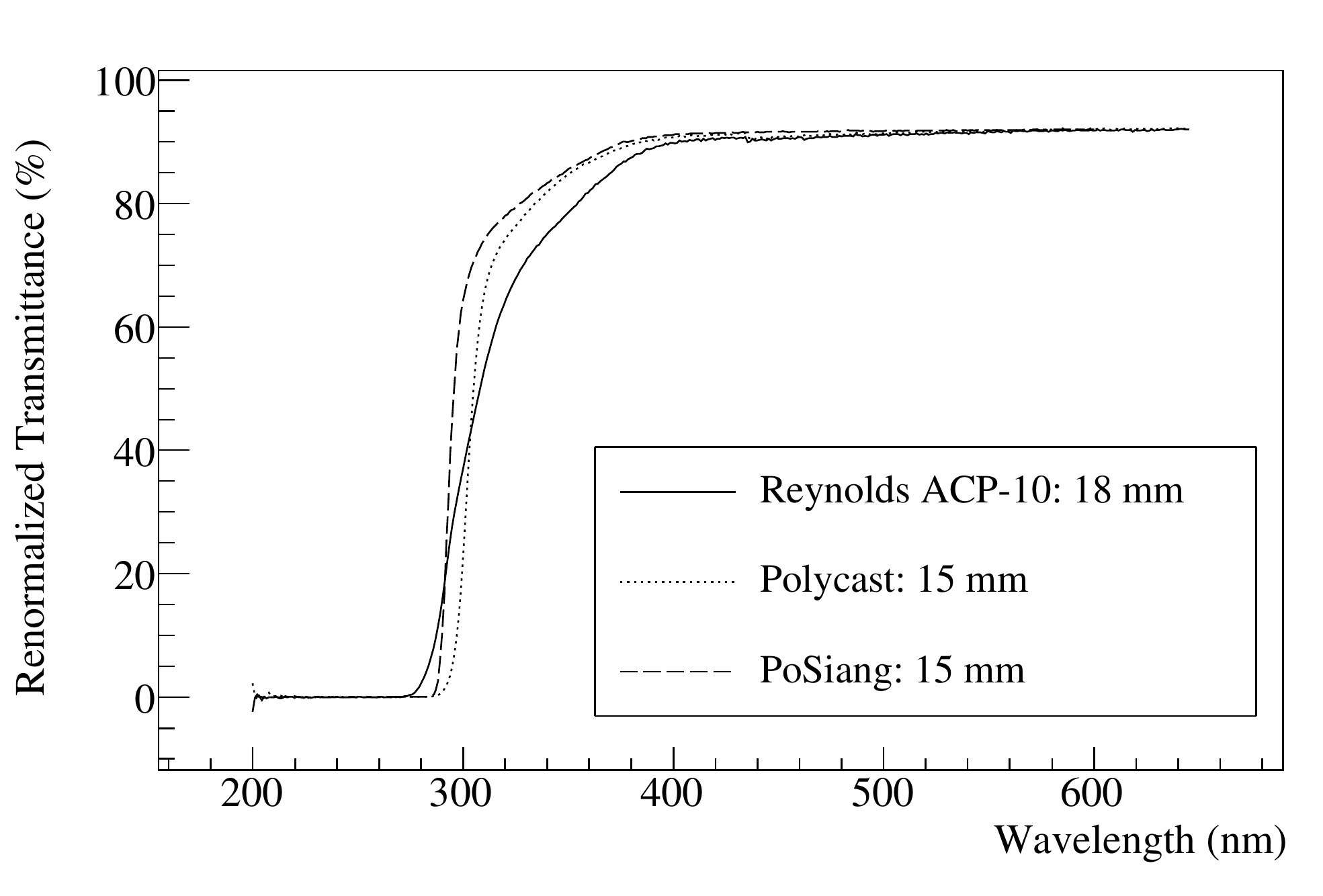}
\caption{Typical transmittances for PoSiang, Polycast, and Reynolds
  UVT acrylics.  Sample pathlengths are given in the legend.  To correct for light collection systematics, Spectra were normalized to give 92\% transmittance at 650~nm, as would be expected for perfect transmittance from a parallel-plane acrylic sample in air.}
\label{fig:Trans}
\end{figure}

Transmittances were measured for each individual production acrylic sheet used in all IAVs and OAVs.  This was done to ensure that poorly transmitting sheets were not included in the vessels, which would have caused asymmetries in detector response.  OAV acrylic sheet transmittance measurements were done using an SI Photonics Model 440 UV-Vis Spectrometer and a Cary Model 300 UV-Vis Spectrometer.  The transmittance of the OAV bonds was also measured using the same equipment.  IAV acrylic sheet transmittance measurements were done using a PerkinElmer Lambda 650 UV-Vis Spectrometer.  A typical transmittance spectrum for the Reynolds and Polycast acrylics used in the OAV and PoSiang acrylic used in the IAV can be seen in Figure~\ref{fig:Trans}.  It was found that all supplied sheets and bonds were optically acceptable for use in the Daya Bay experiment.  The systematic uncertainties in the transmittance measurements are of the order of 1\%.  Uncertainties arise largely from geometric differences between acrylic samples, as well as from the precision of the spectrometers.

\subsubsection{Index of Refraction and Attenuation Length}

Attenuation length and index of refraction are intrinsic material properties, as opposed to transmittance, which is thickness-dependent.  Thus, in order to properly model the Daya Bay detectors, the acrylics' indices of refraction and attenuation lengths must be known.

Light transmission through acrylic and reflection at the acrylic surface are determined by the Fresnel equation and by the Beer-Lambert law, respectively,
\begin{equation}
\label{eq:refl}
R=\frac{(n-1)^2}{(n+1)^2}
\end{equation}
\begin{equation}
\label{eq:trans}
T=e^{-x/\Gamma},
\end{equation}
where n is index of refraction, x is the material's pathlength, and $\Gamma$ is the attenuation length of the material.  In the case of parallel-plane interfaces, such as the acrylic vessel walls, the picture is complicated by multiple internal reflections.  In these cases, the measured transmittance and reflectance of a sample with parallel edges are
\begin{equation}
\label{eq:meastrans}
T^* = \frac{(1-R)T}{1-R^2T^2}
\end{equation}
\begin{equation}
\label{eq:measrefl}
R^* = R(1+TT^*),
\end{equation}
with T and R as given in Equations~\ref{eq:refl} and~\ref{eq:trans}.  By measuring T$^*$ and R$^*$, one can numerically solve for T and R.  This measurement method is called an RT measurement, and has been used to determine the optical properties of UVT acrylics in the past \cite{Zwinkels-1990}.

\begin{table}[htp]
\centering
\begin{tabular}{|c|c|c|c|}
\hline 
Acrylic Type & Wavelength (nm) & RT-Measured n & Schott-Measured n \\ \hline \hline
PoSiang & 365 & 1.512  & - \\ \hline
PoSiang & 405 & 1.504  & - \\ \hline
PoSiang & 480 & 1.495  & 1.4966 \\ \hline
Polycast 1 & 480 & - & 1.4970  \\ \hline
Polycast 2 & 480 & - & 1.4974  \\ \hline
Reynolds & 480 & - & 1.4973  \\ \hline
PoSiang & 546 & 1.491  & 1.4922  \\ \hline
Polycast 1 & 546 & - & 1.4926  \\ \hline
Polycast 2 & 546 & - & 1.4929  \\ \hline
Reynolds & 546 & - & 1.4928  \\ \hline
PoSiang & 587 & 1.490 & 1.4902 \\ \hline
Polycast 1 & 587 & -  & 1.4906  \\ \hline
Polycast 2 & 587 & -  & 1.4910  \\ \hline
Reynolds & 587 & -  & 1.4909 \\ \hline
\hline
\end{tabular}
\caption{Index of refraction measurements using the RT method and measurements procured from Schott NA.  Systematic uncertainties on measurements are 0.006 and <0.001, respectively.}
\label{tab:IOR}
\end{table}

R* and T* measurements for one sample of PoSiang acrylic were done using a Perkin-Elmer Lambda 650 with a 60~mm integrating sphere, which could be set to measure reflectance or transmittance.  The 0.006 uncertainty on the index of refraction measurement was imposed by the systematic uncertainties in the T* and R* measurements.  Results for particular wavelengths can be seen in Table~\ref{tab:IOR}.

As a cross-check, another sample from another sheet of PoSiang acrylic was sent, along with two Polycast samples and one Reynolds sample, to have their indices of refraction measured with a V-block refractometer by Schott North America~\cite{Schott}.  This measurement could be performed at wavelengths above 465~nm with an uncertainty of <0.001.  One can see good agreement between the RT and Schott measurements.  In addition, the low-wavelength RT measurements were in good agreement with low-wavelength RT measurements made in~\cite{Zwinkels-1990}.

It was also noted that all samples displayed very similar indices of refraction: the variation in R resulting from differences in index of refraction of this level are negligible, on the order of 10$^{-7}$.  Given the non-variability of the index of refraction in all acrylics in the study, we treat all acrylics as having identical indices of refraction for the purposes of the Daya Bay Experiment.

\begin{figure}[htp]
\centering
\includegraphics[width=5 in]{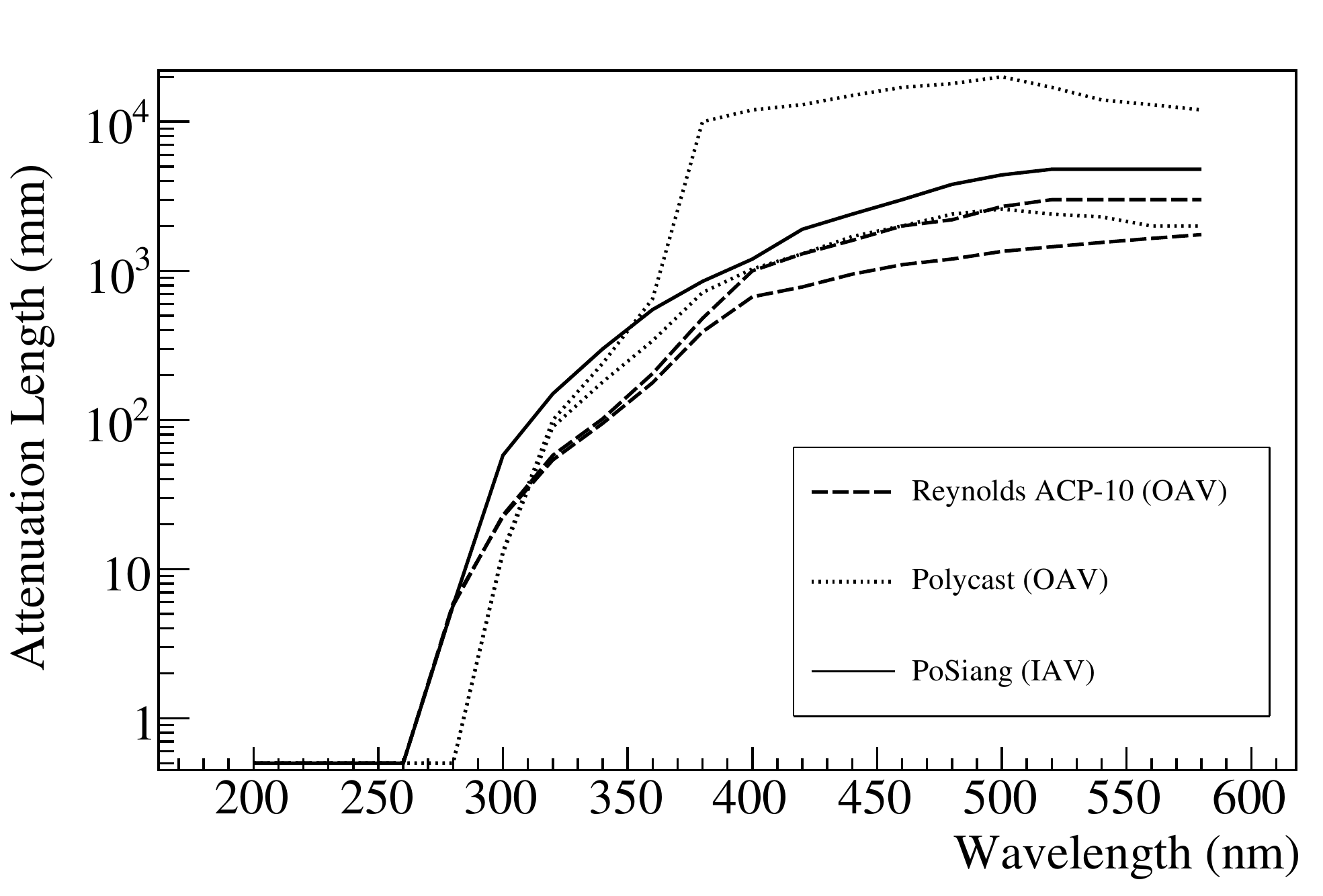}
\caption{Best and worst case attenuation length spectra for each type of IAV or OAV acrylic.  The best and worst values are the product of the transmittance measurement uncertainties as well as actual variations in acrylic attenuation length.  For PoSiang, only one sample's transmittance is pictured.}
\label{fig:att}
\end{figure}

With the index of refraction spectrum determined, attenuation lengths for any individual sheet were calculated using the transmittance measurement done for optical QA.  Attenuation length was calculated for the best- and worst- performing sheets from the transmittance testing, to establish the range of attenuation lengths for all acrylic types.  These spectra can be seen in Figure~\ref{fig:att}.  It should be noted that the difference between the best and worst performing samples for each acrylic type are at least partially caused by the systematic uncertainties from the transmittance measurement.

Acrylic sheet selection and placement in the OAVs and IAVs was uncorrelated with sample optical performace, ensuring no systematic detector-to-detector variations in light yield would arise from this source.  Monte Carlo simulations have shown that the average acrylic pathlength of optical photons in the Daya Bay detector is a few~cm.  This means that the presented range of attenuation lengths would result in percent-level or lower position-related light yield asymmetries between detectors, and negligible overall light yield difference between detectors.

\subsubsection{Ultraviolet Degradation}

It was been demonstrated that UVT acrylics optically degrade when exposed to high levels of UV light \cite{Bryce-2009}.  If such degradation takes place in the AVs during or after construction through exposure to sunlight or excessive factory lighting, the loss of transmittance would not be measured in the QA program, resulting in an unexpected decrease in photoelectron yield during detector operation.  In addition, UV blacklights used to cure acrylic bonds on the IAVs can also be sources of optical degradation.  To quantify these issues, the rates of degradation of the three types of OAV and IAV acrylics were measured.  Based on these rates, safeguards to avoid dangerous UV exposures were implemented.

\begin{figure}[htp]
\centering
\includegraphics[width=5.5 in]{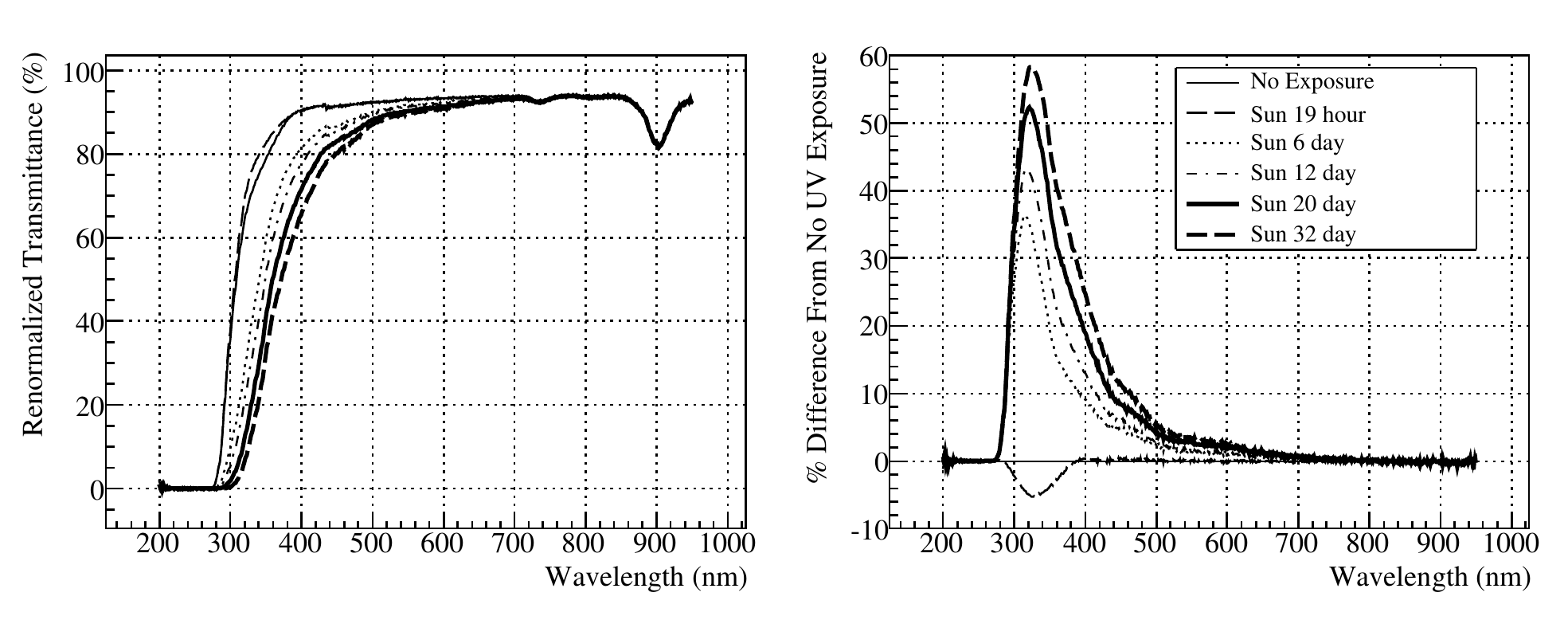}
\caption{Degradation of Reynolds acrylic over time.  Here, one can see in particular which wavelength ranges are most affected by UV degradation.}
\label{fig:Deg}
\end{figure}

The degradation QA program consists of taking two samples of each type of acrylic used in the IAVs and OAVs and subjecting one of each type to either direct sunlight for a period of 30 days, or to UV blacklight-blue light for a period of 30 days.  The transmittance of each sample is measured at numerous times throughout the course of the exposure; the measurements for Reynolds acrylic can be seen in Figure~\ref{fig:Deg}.  At long timescales, degradation effects are very apparent.  One can also see UV curing effects where transmittance initially improves at short wavelengths.  These curing effects dominate in the UV blacklight samples, and no long-term degradation is visible.  Since curing effects are only at low wavelengths, where the Daya Bay liquids emit little light, UV bond curing has negligible effects on light yield for the Daya Bay detectors, and thus likely acceptable practice for most liquid scintillation experiments ultilizing acrylic.

\begin{figure}[htp]
\centering
\includegraphics[width=4 in]{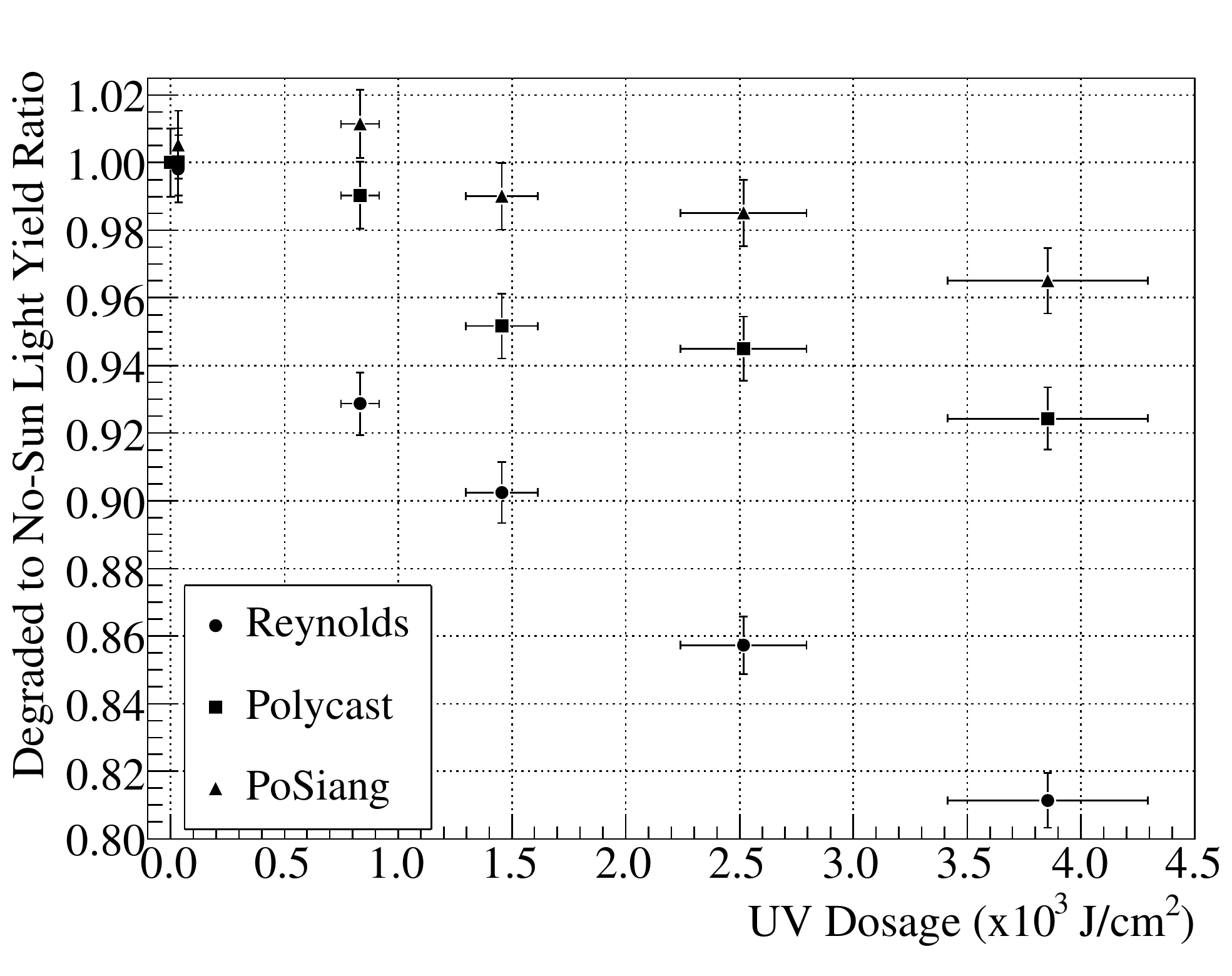}
\caption{Estimated transmittance over the entire wavelength spectrum versus UV dosage for all three types of OAV and IAV acrylic.  The degradation rate is represented by the slope of a line fitted to these points.}
\label{fig:DegRate}
\end{figure}

Using UV meters and weather monitoring websites, total UV dosage can be calculated for each period of sunlight exposure.  Then, by combining the AD liquid emission spectrum with the measured transmittances, a relationship between UV dosage and light yield loss can be established, as in Figure~\ref{fig:DegRate}, for sun exposure.  By fitting these points one can procure the decay rates of the acrylic and calculate acceptable UV exposure times for each component based on the expected average acrylic pathlength of optical photons in the Daya Bay detector.  Acceptable exposure time is defined as the amount of time in which UV exposure caused a 2\% decrease in total estimated light yield from one vessel.  These values for each component are listed in Table~\ref{tab:ExpTimes}. 

\begin{table}[htp]
\centering
\begin{tabular}{|c|c|c|c|}
\hline 
Acrylic Part & Sun Exposure Limit (d) & Factory Limit (d) & UV-Filtered Factory Limit (d) \\ \hline \hline
OAV Lid and Base & 0.5 & 16  & 205 \\ \hline
OAV Barrel & 1.2 & 40  & 527 \\ \hline
IAV  & 3.5  & 122 & 1590 \\ \hline
\end{tabular}
\caption{Acceptable exposure times for the OAV and IAV in various environments.  The factory limits are quoted with and without UV-filtered windows.}
\label{tab:ExpTimes}
\end{table}

Exposure times are kept underneath these values by altering shipping and storage procedures accordingly.  Acrylic components are kept under plastic wrap or tarps while not being directly worked on.  The vessels or individual sheets are never kept outside uncovered.  UV filters have also been installed on windows of the buildings in which the vessels were manufactured, stored, cleaned or installed.

\subsubsection{Acrylic Vessel Stress Testing}

As mentioned before, the specification for maximum acceptable long-term stress on any area of the AV is 5~MPa.  While Reynolds and Nakano annealed the vessels to reduce residual stress levels, acrylic stress measurements were not performed by Reynolds on the OAVs. To evaluate residual stresses, a method for quantitatively measuring stress on the constructed vessels using a StrainOptics PS- 100-LF-PL polarimetry system was developed \cite{Strainoptics}.

\begin{figure}[htpb]
\centering
\subfigure{
  \includegraphics[width=0.6\textwidth]{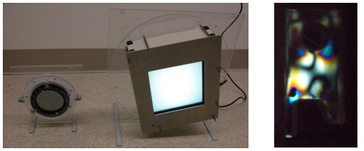}
  \label{subfig:StressDevice}
}
\hspace{0.5cm}
\subfigure{
  \includegraphics[width=0.3\textwidth]{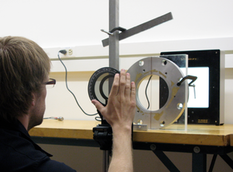}
  \label{subfig:StressPolar}
}
\caption{Photographs of the two components of the stress-measurement system: the larger illuminator and smaller analyzer.  Also pictured is a demonstration of the system's use, as well as the effect of its use: birefringence made visible in a OAV prototype lifting tab resulting from residual stresses.}
\label{fig:StressMeasure}
\end{figure}

The system utilizes the change in index of refraction with introduced stress experienced by acrylic to retard components of polarized light in stressed acrylic regions.  This creates interference patterns, which can be quantitatively linked to stress values using a system of rotating quarter wave plates. The entire system needs to be rotated into alignment with the principal stress axis of points of suspected high stress, and thus a rotating frame was developed for both system components. The two system components can be seen in Figure~\ref{fig:StressMeasure}, along with a fringe pattern exhibited by residual fabrication stresses in the OAV lifting hook.  

The stress measurement system was calibrated using the acrylic stressing setup described in Section 6.1.4.  The device was then brought to Reynolds to be used during QA testing of the prototype and production OAVs.  Stress-testing results are listed in Table~\ref{tab:stresslevels}.  The uncertainty in the provided measurements are dependent on the thickness of the acrylic sample, but it most cases is less than 10\%.  All measured values are below the 5~MPa long-term maximum stress limits.  Qualitative stress testing was performed on the IAVs by rotating polarized filters on either side of an acrylic sheet to look for areas of high birefringence.

\begin{table}[htp]
\centering
\begin{tabular}{|c|c|}
\hline
Location on OAV & Highest Measured Stress (MPa) \\ \hline \hline
Lifting Tabs & 1.3 \\ \hline
Bonds & 3 \\ \hline
Crazed areas along equator bond & $\sim$0 \\ \hline
Lid ribs with lid unattached to OAV & 3 \\ \hline
\end{tabular}
\caption{Stress levels measured in areas of concern for OAV1 and OAV2. The highest measured stress in each location on either OAV is quoted here.}
\label{tab:stresslevels}
\end{table}

The main drawback of this measurement system is that it measures integrated stress through the entire thickness of an acrylic part, making it impossible to measure stresses on individual surfaces in some parts of the OAVs.  For example, the slightly crazed areas near the OAV equator bond, pictured in Figure~\ref{fig:Bond}, despite being annealed, should exhibit residual stress from the outward force applied to hold dams around the equator during the bonding process.  However, this type of force may exert a tension stress on the OAV outside and a compression stress on the OAV inside in that area, which cause birefringence in opposite directions.  The net effect is that the compression and tension integrate through the piece as the polarized light traverses the wall, resulting in no net observed birefringence, despite the existence of stresses.  This is the case in Figure~\ref{fig:Bond}; microcrazing is present in the black area above the visible bond line, and no microcrazing was identified below the bond line.  Yet, both areas are exhibit the same lack of birefringence.  This drawback does not affect the validity of the measurements on the lifting tabs, bonds, or lid ribs, as the stresses experienced in these areas would be mostly uniform through the entire length of acrylic traversed by the polarized light.

\subsection{Geometric Characterization}

The geometric parameters of the Daya Bay AVs must be measured to ensure identicalness of detector shape and target mass between detectors, and to make sure the detector will fit together properly during assembly.  Dimensional parameters were measured at either Reynolds or Nakano, while the mass and survey data were made at the Daya Bay SAB.

\subsubsection{Wall thickness}

\begin{figure}[htpb]
\centering
\subfigure[OAV thickness deviation with height on the OAV.  Measurements of various angles were aggregated to make each data point.] {
  \includegraphics[width=0.45\textwidth]{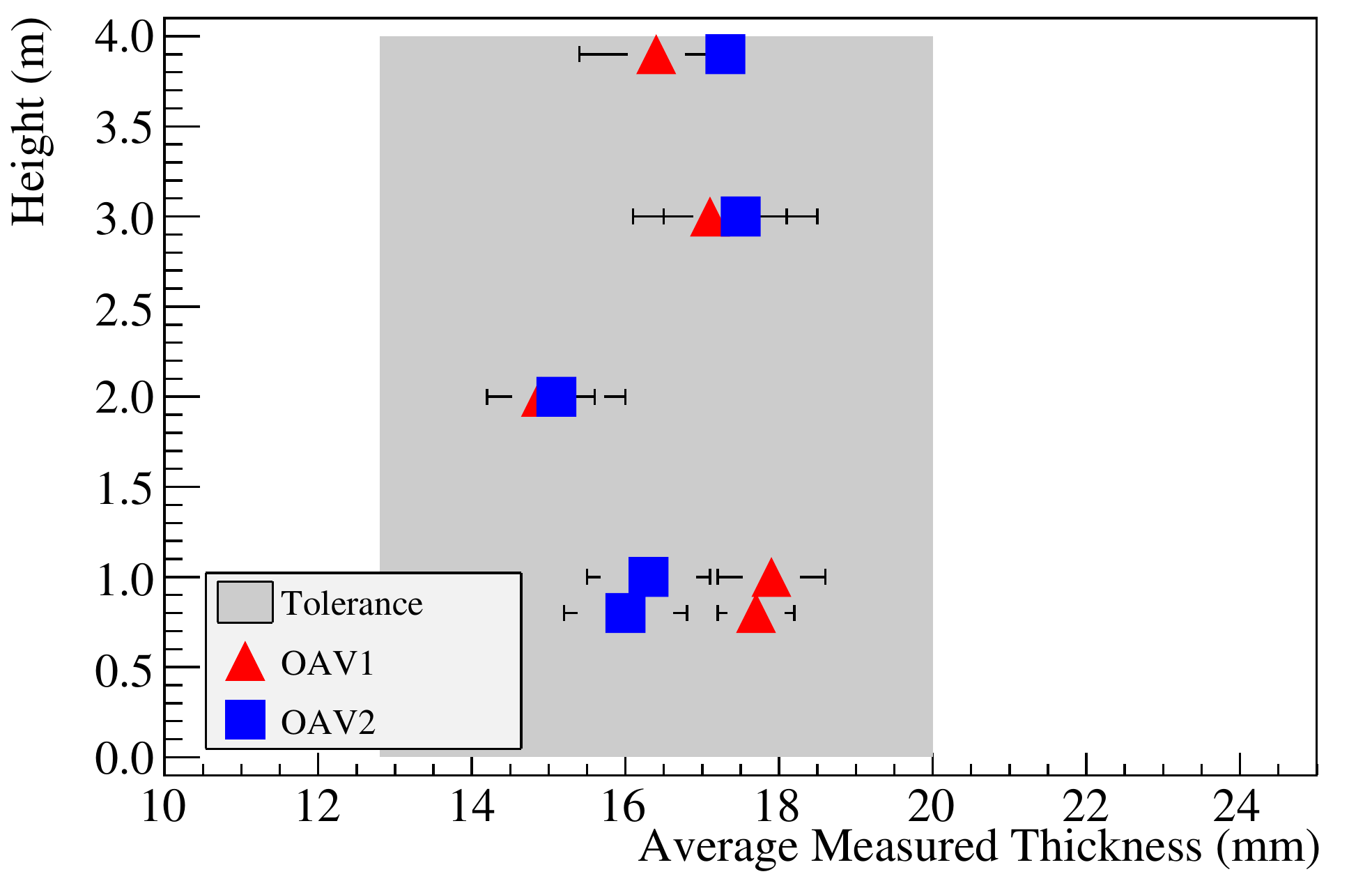}
  \label{subfig:OAVThick}
}
\hspace{0.5cm}
\subfigure[IAV thickness deviation with angular position around the IAV edge.  Measurements of various heights were aggregated to make each data point.] {
  \includegraphics[width=0.45\textwidth]{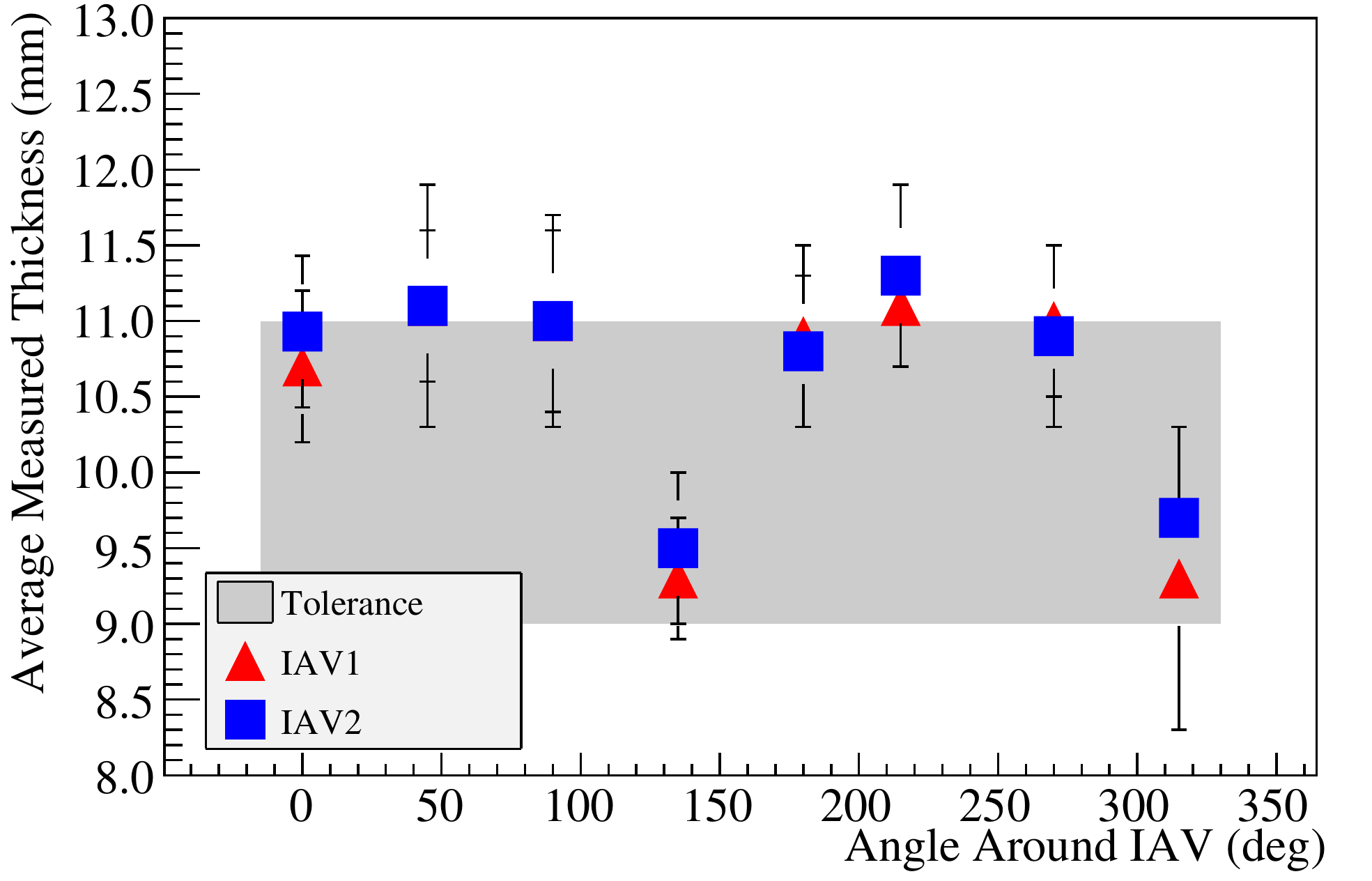}
  \label{subfig:IAVThick}
}
\caption{IAV and OAV thickness measurements for both vessels, measured at various AV heights and angles.  The error bars indicate the standard deviation of all thickness measurements made at the given height or angle. Nearly all measurements are within the tolerances indicated by the grey band in each figure.}
\label{fig:thickness}
\end{figure}

The designed thickness of the OAV barrel, top, and bottom is 18~$\pm$5.2~mm; those values for the IAV are 10~$\pm$1~mm for the barrel, and 15~$\pm$1~mm for the bottom and top.  Barrel wall thickness measurements are made at various heights and angles around the vessel with an ultrasonic thickness gauge.  Data for IAV and OAV barrels can be seen in Figure~\ref{fig:thickness}.

The average vessel wall thickness was 16.8~$\pm$1.3~mm and 16.4~$\pm$1.2~mm for OAV1 and OAV2 and 10.55~$\pm$0.98~mm and 10.69~$\pm$0.8~mm for IAV and IAV2.  These values are within the acceptable thickness range.  There were a few isolated areas, mostly along bonds, where thickness was below tolerance by 1-2 mm; these areas accounted for less than 1\% of the total barrel surface area, and are not a concern.  Vessel thinness in bond areas likely originates in the sanding and polishing process.  When sheets are bonded together and polished, any surface discontinuities from uneven sheet alignment or shape are evened out, which, in some cases, results in the removal of a significant amount of acrylic material.  Less systematic checks of OAV and IAV top and bottom thickness were also done, showing no deviations from tolerances.

\subsubsection{Diameter}

\begin{figure}[htpb]
\centering
\subfigure[Measured OAV diameter variations with height. A small but clear difference in diameter is visible.] {
  \includegraphics[width=0.45\textwidth]{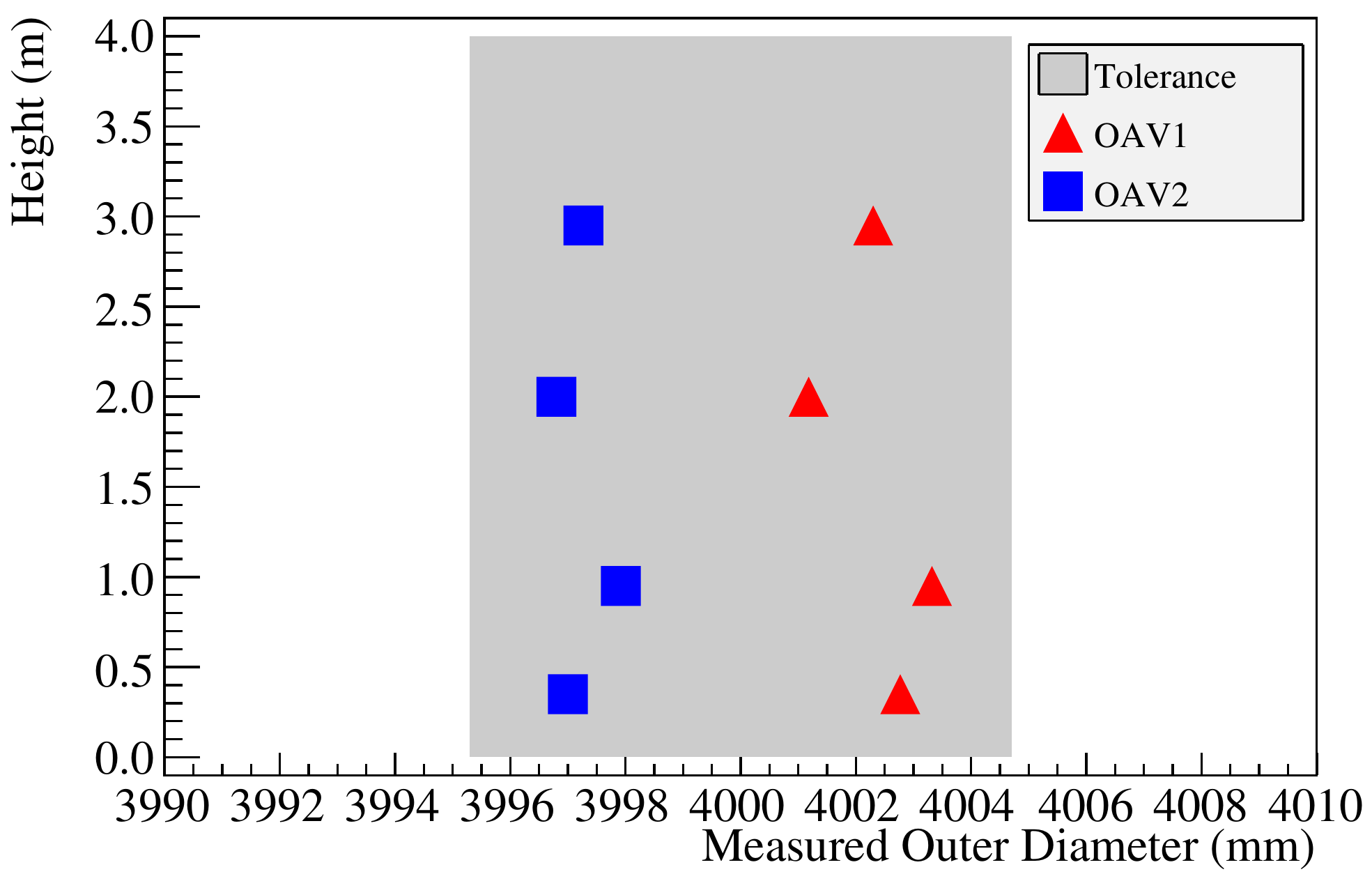}
  \label{subfig:OAVDiam}
}
\hspace{0.5cm}
\subfigure[Measured IAV diameter variations with height.] {
  \includegraphics[width=0.45\textwidth]{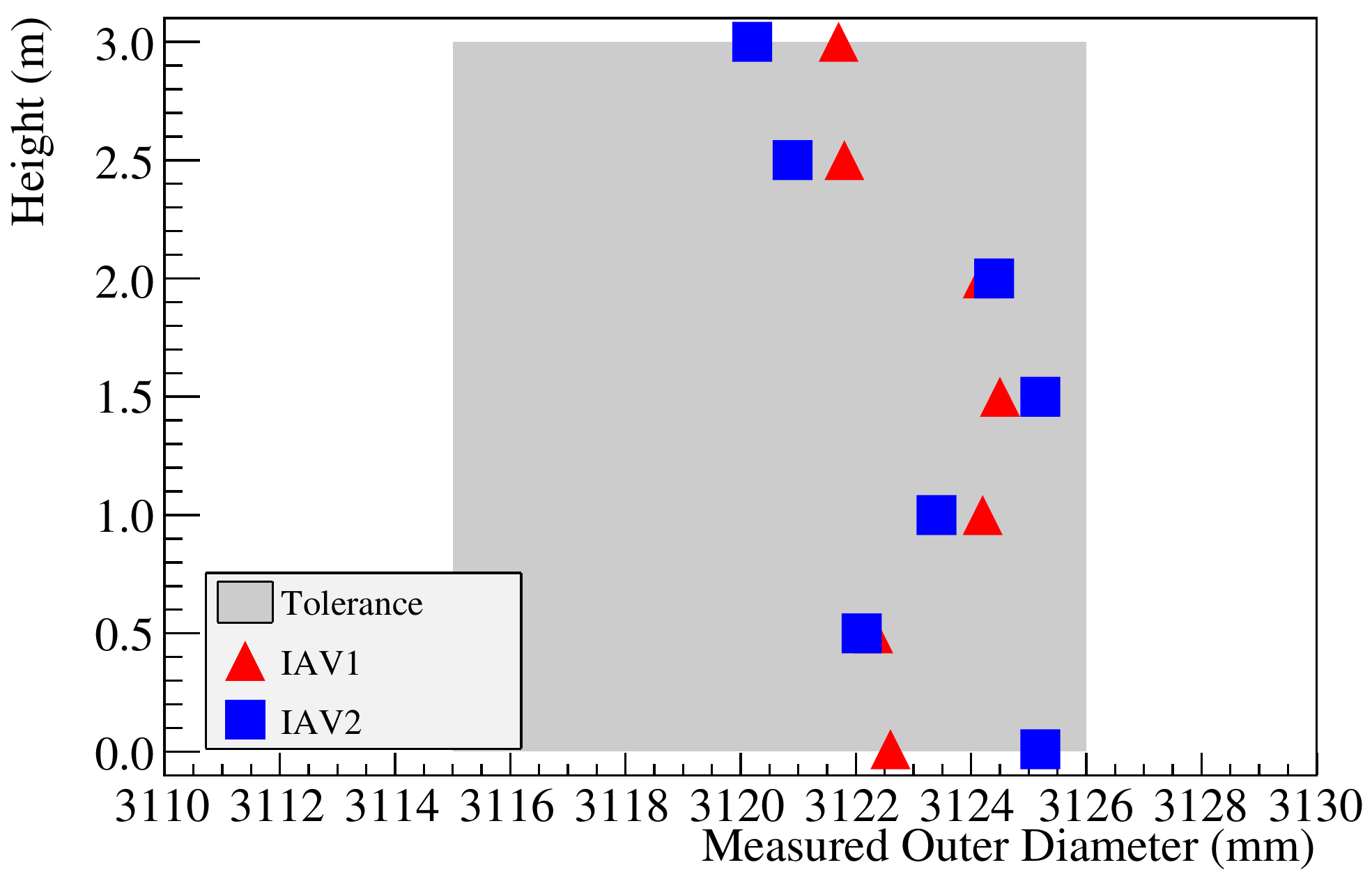}
  \label{subfig:IAVDiam}
}
\caption{IAV and OAV diameter measurements for both vessels, measured at various heights on the AVs.  Only one measurement was taken at each height.  Tolerances are indicated by the grey band in each figure.}
\label{fig:diam}
\end{figure}

Designed outer diameters are 4000~$\pm$4.8~mm for the OAVs and 3120~$\pm$5~mm for the IAVs.  Outer diameters were measured at various heights on all four AVs using a pi tape measure.  Outer-diameter data for all four AVs can be seen in Figure~\ref{fig:diam}.  Average diameter was 4002.4~$\pm$0.9~mm and 3997.4~$\pm$0.5~mm for OAV1 and OAV2 and 3123.1~$\pm$1.2~mm and 3123.1~$\pm$2.0~mm for IAV1 and IAV2.  The systematic 0.5~cm diameter difference between OAVs is within engineering tolerance, and will not have a significant impact on detector response.

For the OAVs, inner diameter was also measured to ensure that the vessels had a sufficiently circular cross-section.  The specified inner diameter is 3964~$\pm$13.6~mm, and was measured with a Leica laser distance measure.  Average values were 3968.6~$\pm$7.8~mm and 3965.3~$\pm$5.04~mm for OAV1 and OAV2, respectively.  These values match well with the outer diameter measurements in showing that OAV1 is slightly larger than OAV2.  The large majority of measurements were well within tolerance.



The inner vessels were not designed to be entered and exited, so inner diameter measurements were not made.  Instead, the horizontal cross-section of the vessel was checked by measuring angles around the outside of the IAV across the bottom inside of the IAV.  The angle between two positions across the AV from one another as measured from a location halfway between those points along the AV's circumference was measured to be between 89.8$^{\circ}$ and 90.1$^{\circ}$, well within the specified range of 89.2$^{\circ}$ and 90.8$^{\circ}$.

\subsubsection{Height}

\begin{figure}[htpb]
\centering
\subfigure[Measured OAV height variations with angle around the OAV outside edge.] {
  \includegraphics[width=0.45\textwidth]{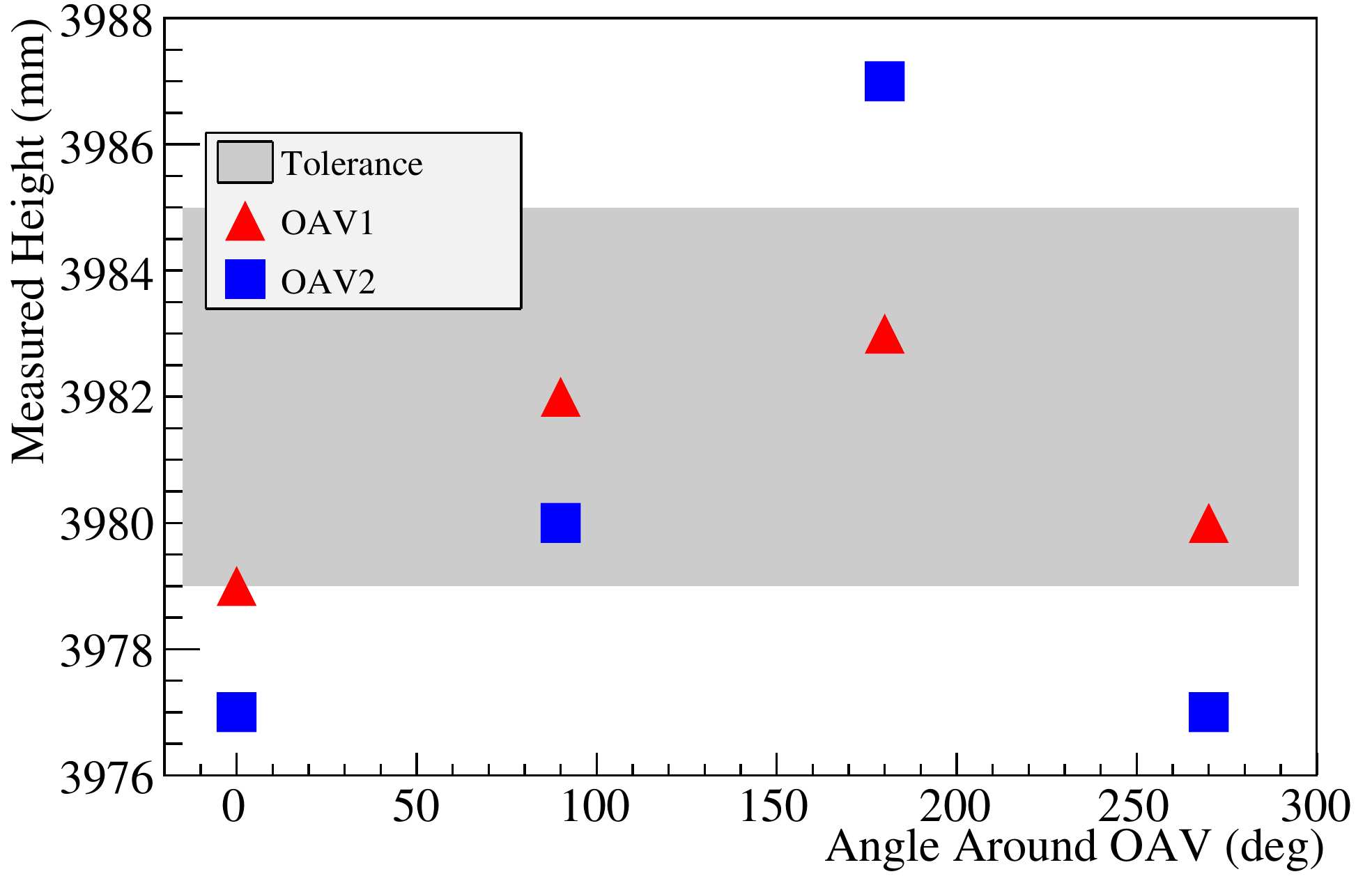}
  \label{subfig:OAVHeight}
}
\hspace{0.5cm}
\subfigure[Measured IAV height variations with angle around the IAV outside edge.] {
  \includegraphics[width=0.45\textwidth]{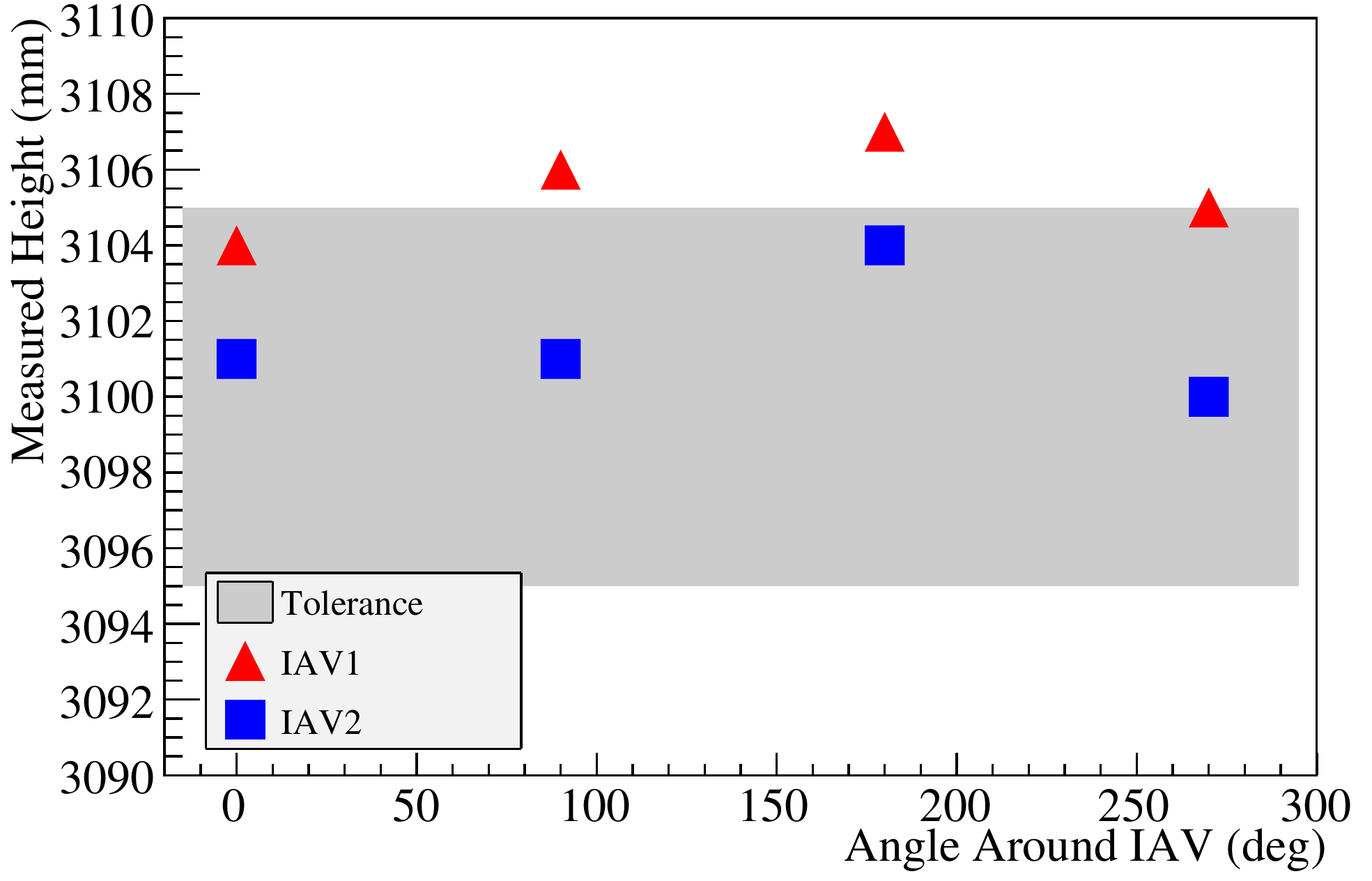}
  \label{subfig:IAVHeight}
}
\caption{IAV and OAV height measurements for both vessels, measured at various angles on the AVs.  Only one measurement was taken at each height.  Tolerances are indicated by the grey band in each figure.}
\label{fig:height}
\end{figure}

Specified heights, defined as the distance from the AV bottom to AV flange for the OAV and from the AV bottom to AV top edge for the IAV, are 3982~$\pm$3.0~mm for the OAVs and 3100~$\pm$5.0~mm for the IAVs.  Heights were measured on all vessels with a tape measure, and can be seen in Figure~\ref{fig:height}.  The average height values were 3981.0~$\pm$1.8~mm and 3980.25~$\pm$4.7~mm for OAV1 and OAV2 and 3106~$\pm$1.31~mm and 3101~$\pm$1.55~mm for IAV1 and IAV2.  One can see that IAV1 is on average taller than IAV2 by 0.5~cm.  Some of OAV2's heights were out of spec by a few mm, but this difference was deemed acceptable for the OAVs.

\subsubsection{Perpendicularity and Profile}

To test for straightness of the OAV barrel as well as the barrel's perpendicularity with respect to the bottom of the OAV, a plumb bob test was done.  First, a plumb bob was hung from a string taped to the top of the OAV flange.  The distance between the string and vessel walls was then measured.  The specification on the overhang of the OAV flange over the outside of the OAV walls was 40~$\pm$10~mm.  Results from OAV1 can be seen in Table~\ref{tab:bob}.

\begin{table}[htp]
\centering
\begin{tabular}{|l|l||c|c|c|c|c|c|c|}
\hline
\multirow{2}{*}{Height (m)}  & \multicolumn{8}{|c|}{Angle (deg)} \\ \cline{2-9}
 & 0 & 45 & 90 & 135 & 180 & 225& 270 & 315 \\ \hline
\hline
 0.35 & 51 & 43 & 43.5 & 45.5 & 33.5 & 32 & 40 & 41  \\ \hline
 1.0 & 49 & 40 & 43.5 & 43.5 & 32 & 32 & 43.5 & 40 \\ \hline
 2.0 & 50 & 40 & 45 & 45.5 & 33 & 33 & 50.5 & 40 \\ \hline
 3.0 & 46 & 41 & 44 & 44 & 37 & 38 & 46 & 41 \\ \hline
 3.9 & 43.5 & 42 & 42 & 43.5 & 40 & 42.5 & 42 & 42 \\ \hline
\end{tabular}
\caption{Measured distance between a plumb bob string and the side of OAV1 in mm.  The specification on this value is 40~$\pm$10~mm.}
\label{tab:bob}
\end{table}

The proper orientation of the bottom and the proper wall shape is reflected in the plumb bob data: almost all measured vessels were within the specification.  Minor wall features are described by the data: for example, at 270$^{\circ}$, the vessel bulges outward in the middle by about 1~cm, while at other angles, like 180$^{\circ}$, the vessel middle bulges inward.  None of these features are expected to affect appreciably the physics response or mechanical stability of the detector.  The measurements from OAV2 are also similarly within tolerance.

\subsubsection{O-Ring Grooves}

To make sure that the vessels do not leak, dimensional measurements of the OAV o-ring groove were carried out separately from the standard leak-checking regimen.  If the groove is too deep, there will not be enough pressure on the o-ring to seal properly, while if it is too shallow, the o-ring will be overcompressed and may lose its elasticity.  

The height and width of the two concentric OAV flange o-ring grooves need to meet specifications, which are 4.4$\pm$0.3~mm and 6.9$\pm$0.3~mm, respectively.  O-ring groove dimensions were measured at various angles with calipers accurate to 0.1~mm.  Average o-ring groove width was 6.7$\pm$0.2~mm and 6.8$\pm$0.3~mm for OAV1 and OAV2, respectively.  While widths were in tolerance, o-ring groove depths were initially too shallow, caused by the removal of excess material on the OAV flange during sanding and polishing.  This was fixed by Reynolds before shipment by making a specially-designed sanding tool to remove more acrylic from the bottom of the grooves.  The grooves were then remeasured at an average depth of 4.4$\pm$0.3~mm.  O-ring grooves were cut deeper on future OAVs to compensate for the loss of flange material during polishing.



\subsubsection{Vessel Positioning: In-Situ Survey}

During installation, position surveys are done on all major AD components to assure that they are placed correctly in the AD.  The surveys are done using a Leica System 1200 Total Station.  Any group of survey measurements are started by placing the Total Station and tripod at a fixed location in the cleanroom.  The coordinate system is set by taking the coordinates of a fixed reference point.  The Total Station is then aimed at retroreflectors secured to the desired survey locations.  The Total Station automatically locates the retroreflector using a laser sight and records x- y- and z-coordinates.  Using this method, surveys were done of points on the SSV and OAV bottom, sides, and lid, and on the IAV top.

\begin{figure}[htb]
\centering
\includegraphics[trim=2cm 7cm 4cm 2cm, clip=true, width=5 in]{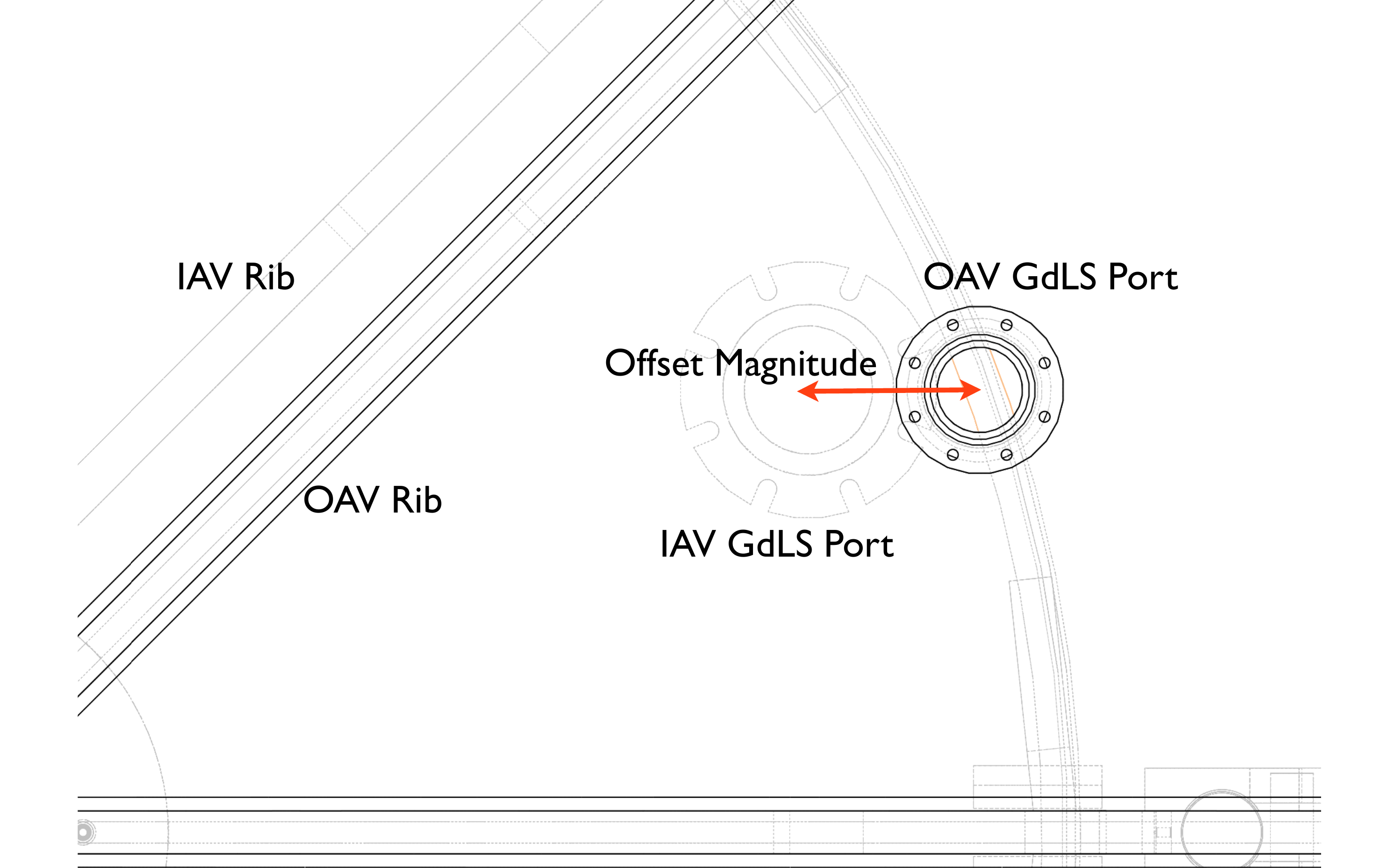}
\caption{Top view engineering drawing showing exaggerated offsets between IAV and OAV ports.  The offset pictured here is made greater than the actual measured offsets so that misalignment of the ports is easily visible.}
\label{fig:ports}
\end{figure}

\begin{figure}[htb]
\centering
\includegraphics[width=5.5 in]{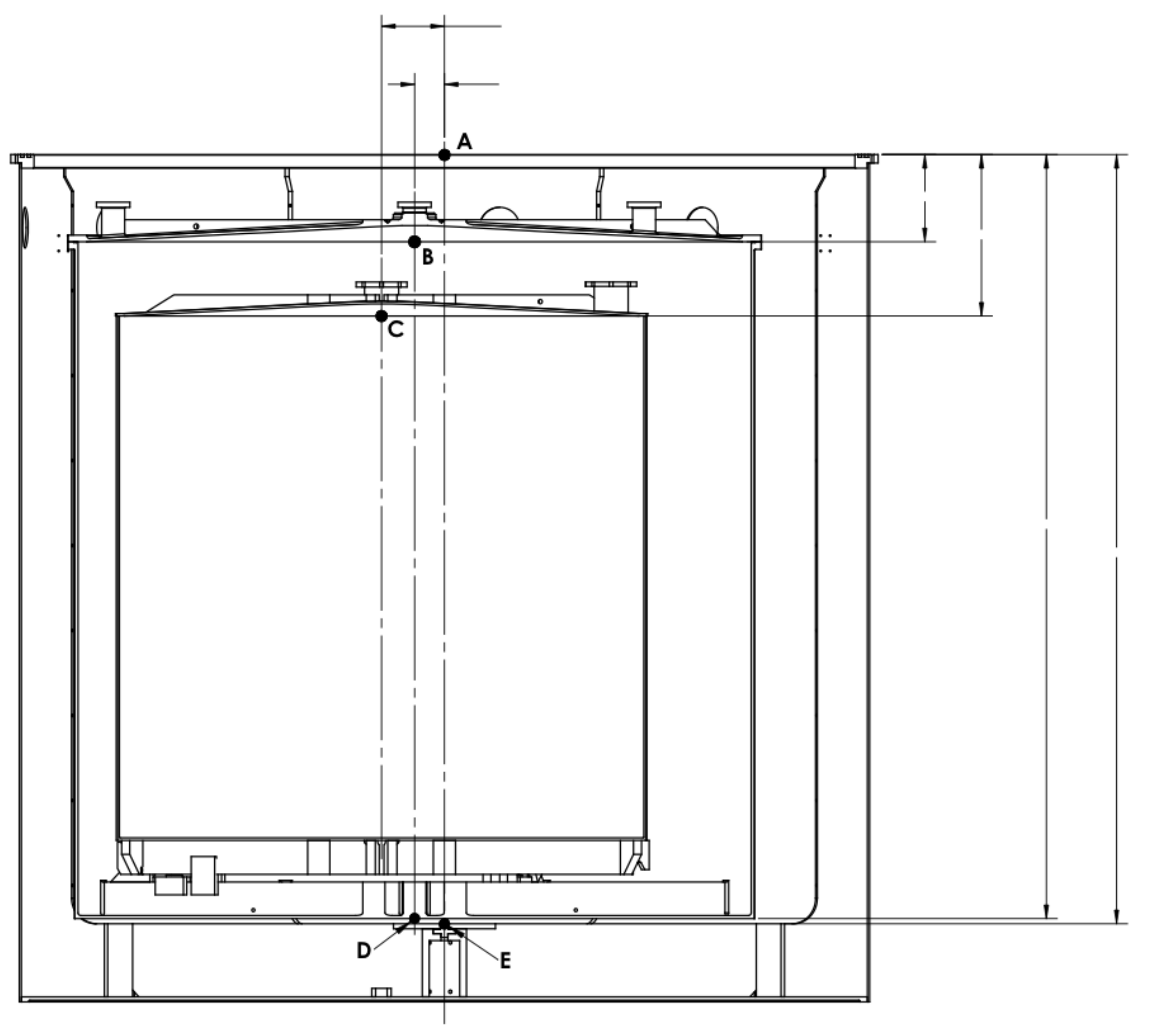}
\caption{Depiction of an AD with offset AVs that shows key reference points (B, C, D, and E) whose vertical and horizontal distances from the calculated SSV lid center (point A) are calculated from the AV survey data.}
\label{fig:dims}
\end{figure}

The azimuthal orientations of the OAVs and IAVs are dictated by the alignment of the SSV, OAV, and IAV calibration ports.  The flexibility of the bellows connecting the calibration and overflow systems to the AVs allows a small amount of acceptable misalignment.  The port location survey data is pictured in Table~\ref{tab:portalign}.  The dimensions presented in this table are illustrated in Figures~\ref{fig:dims}~and~\ref{fig:ports}.

\begin{table}[htp]
\centering
\begin{tabular}{|c|c|c|c|c|c|c|}
\hline
\multirow{2}{*}{Vessel} & \multicolumn{2}{|c|}{Central Port} & \multicolumn{2}{|c|}{Gd-LS Calib. Port} & \multicolumn{2}{|c|}{LS Calib. Port} \\ \cline{2-7}
&Magnitude (mm) & Z (mm) & Magnitude (mm) & Z (mm) & Magnitude (mm) & Z (mm) \\ \hline \hline
SSV1 & 0 & 19 & 0 & 1 & 0 & 3 \\ \hline
SSV2 & 0 & 32 & 0 & 5 & 0 & 6 \\ \hline
OAV1 & 2.6 & 321 & 20.4 & 312 & 3.6 & 313 \\ \hline
OAV2 & 2.3 & 324 & 5.0 & 312 & 3.2 & 312 \\ \hline 
IAV1 & 2.0 & 783 & 4.7 & 789 & - & - \\ \hline
IAV2 & 2.6 & 792 & 2.5 & 783 & - & - \\ \hline
\end{tabular}
\caption{Port alignment data.  The magnitudes of x-y plane offsets for AVs are given with respect to the corresponding SSV ports.  The z-positions are all given with respect to the top of the SSV lid center.}
\label{tab:portalign}
\end{table}

All ports are aligned to within 5~mm with the exception of the IAV1 Gd-LS port, which is misaligned by 2~cm.  While not ideal, the connection hardware was able to flex to make up for this larger than expected misalignment.  The Z-positions of the various ports relative to the SSV top are mostly consistent between detectors to within 1~cm.  The discrepancy between SSV central port Z position relative to the SSV top is likely a relic of using different retroreflector stands for each SSV.  It is also interesting to note that the central port of each OAV appears to be around 1~cm lower than that OAV's LS port, indicating deflection of the OAV lid under its own weight. This sag will be reduced when the detector is filled with liquids.

The shapes of the LS and MO regions in the as-built antineutrino detectors are partially determined by the physical positioning of the OAV and IAV.  AV positioning information consists of levelness, Z, and R positions of the IAV, OAV, and SSV, and is provided by survey data gathered by the method described above.  The survey data is analyzed to give the horizontal and vertical positions of the calculated centers of specific surfaces on the AVs and SSV relative to the calculated center of the SSV lid.  These dimensions correspond to the horizontal and vertical distances between point A and points B, C, D, and E in Figure~\ref{fig:dims}.

Table~\ref{tab:concentric} presents an overview of the surveyed vessel offsets between point A and points B, C, D, and E in cylindrical coordinates: magnitude of Z offsets of the centers of given vessel surfaces are presented along with magnitudes of X-Y offsets.  In addition, levelness is described by the difference in absolute Z-coordinate between one side of a vessel and the other.  The vessels are largely concentric, having less than 5~mm offset in most cases.  Z-positions of the various surfaces are in close agreement between ADs, and tilts of these surfaces are all around 5~mm or less.

\begin{table}[htp]
\centering
\begin{tabular}{|c|c|c|c|c|}
\hline
Vessel & Magnitude (mm) & Angle (deg) & Height (mm) & Tilt (mm) \\ \hline \hline
SSV1 Top & - & - & - & 4 \\ \hline
SSV2 Top & - & - & - & 2 \\ \hline
OAV1 Flange & 2 & 257 & 542 & $<$4.4 \\ \hline
OAV2 Flange & 7 & 39 & 546 & $<$4 \\ \hline
IAV1 Lid Edge & 3 & 256 & 970 & 5 \\ \hline
IAV2 Lid Edge & 2 & 259 & 964 & 6 \\ \hline
OAV1 Bottom & 2 & 0 & 4515 & 4 \\ \hline
OAV2 Bottom & 4 & 231 & 4514 & 3 \\ \hline
SSV1 Bottom & 3 & 352 & 4556 & 3 \\ \hline
SSV2 Bottom & 5 & 186 & 4557 & 2 \\ \hline
\end{tabular}
\caption{As-built vessel surface offset data, listed in cylindrical coordinates as X-Y offset magnitude, X-Y offset angle, and Z offset.  The coordinate system is set such that the top of the SSV lid center is (0,0,0).  Values are obtained from aggregating data from numerous survey locations.  Uncertainties on these aggregated values are $\pm$3~mm.}
\label{tab:concentric}
\end{table}

Survey data was also taken to determine how well the IAV position was constrained within an OAV; While the IAV hold-downs constrain the IAV in z and $\theta$, it is the IAV ribs interfacing with the OAV rib stops that constrain the IAV's position in r.  The difference in length between the IAV bottom ribs and the OAV rib stop gap was surveyed and then decreased to an acceptable 3~mm difference by affixing teflon spacers to the IAV ribs ends.

\subsubsection{Vessel Mass}

The most accurate way to determine the total amount of acrylic in the antineutrino detectors is to measure the acrylic vessels' masses.  This data also provides a measure of the identicalness of the acrylic vessels in pairs of detectors.  The mass of each acrylic vessel was measured in between the cleaning of the vessels and their installation in the neutrino detectors in the Daya Bay Surface Assembly Building.  While the vessel was lifted using a 10-ton crane, a Dillon ED-Junior crane scale placed in series between the crane and the acrylic vessel measured the mass of the vessel.  The total measured masses are 1847$\pm$7~kg (OAV1), 1852$\pm$7~kg (OAV2), 907$\pm$7~kg (IAV1), and 916$\pm$7~kg (IAV2).  Within the uncertainties of the measurement the masses are largely consistent with one another, indicating very equal amounts of non-scintillating material in the center regions of the filled ADs.

\subsubsection{Vessel Volume}

By combining a weight measurement with a measurement of the density of each type of acrylic, the total volume of acrylic in each detector can be calculated.  The density of the three different acrylic types were determined by measuring the mass and volume of acrylic samples using a scale and caliper, respectively.  The density of all acrylics used in the construction of the vessels were measured to be 1.19~g/cm$^3$, with an error of 0.01~g/cm$^3$.  Given this value, the total volumes are 1.552$\pm$0.014~m$^3$ (OAV1),  1.558$\pm$0.014~m$^3$ (OAV2),  0.762$\pm$0.009~m$^3$ (IAV1), and  0.770$\pm$0.009~m$^3$ (IAV2).

\subsubsection{Expected Liquid Volumes}

The mass of the Gd-LS and LS will be measured to 0.1\% or better during the filling process using load cells and Coriolis flowmeters.  Prior to the filling of the AD we can make an estimate of the AD inner volumes using the geometric data of the acrylic vessel characterization.  We use the averaged geometric value as the dimension, with the standard deviation of that value as the uncertainty.  These values are listed in Table~\ref{tab:averagedims}.  For the Gd-LS region, we use the measured dimensions of the vessels along with the design value for the conical top of the vessel; we assign an uncertainty of 5\% to this latter value.  We obtain 23.44$\pm$0.027~m$^3$ and 23.40$\pm$0.027~m$^3$ as the nominal inner target vessel volumes for the Gd-LS in AD1 and AD2, respectively.  This value, along with the previously calculated acrylic volume and OAV dimensions, can be used to deduce the LS volumes.  The OAV bottom rib volume was estimated based on the as-designed values, with an uncertainty of 5\%.  The calculated LS volumes are 25.18~$\pm$0.053~m$^3$ and 25.05$\pm$0.076~m$^3$ for the LS in AD1 and AD2, respectively.   The estimated uncertainties of the Gd-LS and LS volume from the geometric characterization before filling are 0.15\% and 0.2-0.3\%, respectively.

\begin{table}[htp]
\centering
\begin{tabular}{|c|c|c|c|c|c|c|c|c|}
\hline
\multirow{2}{*}{Parameter} & \multicolumn{2}{|c|}{IAV1} & \multicolumn{2}{|c|}{IAV2} & \multicolumn{2}{|c|}{OAV1} & \multicolumn{2}{|c|}{OAV2} \\ \cline{2-9}
& Value & Unc. & Value & Unc. & Value & Unc. & Value & Unc. \\ \hline \hline
Thickness (mm) & 10.55 & 0.98 & 10.69 & 0.8 & 16.8 & 1.3 & 16.4 & 1.2 \\ \hline
Diameter (mm) & 3123.1 & 1.2 & 3123.1 & 2.0 & 4002.4 & 0.9 & 3997.4 & 0.5 \\ \hline
Height (mm) & 3106 & 1.31 & 3101 & 1.55 & 3981 & 1.8 & 3980.2 & 4.7 \\ \hline
Mass (kg) & 907 & 7 & 916 & 7 & 1847 & 7 & 1852 & 7 \\ \hline
Volume (m$^3$) & 0.762 & 0.009 & 0.770 & 0.009 & 1.552 & 0.014 & 1.558 & 0.014 \\ \hline
Volume Inside (m$^3$) & 23.44 & 0.027 & 23.40 & 0.033 & 25.18 & 0.053 & 25.05 & 0.76 \\ \hline
\end{tabular}
\caption{Aggregated average values of major detector parameters for AD1 and AD2 AVs.  Uncertainties are quoted along with each parameter.  The volume inside parameter describes the Gd-LS volume for each IAV and the LS volume for each OAV.}
\label{tab:averagedims}
\end{table}

\section{Summary}

We have successfully performed the design, fabrication, characterization, shipment, and installation of the acrylic vessels for the first two antineutrino detectors for the Daya Bay experiment.  Nested pairs of acrylic vessels are a novel approach to the construction of a 3-volume detector filled with scintillating liquids for the precision measurement of reactor antineutrinos.   An extensive production monitoring and quality assurance program ensures that the arylic vessels meet the technical and scientific requirements for the Daya Bay experiment. The nested acrylic vessels exhibit a high degree of identicalness. This is a critical feature of the pair-wise measurement of reactor antineutrinos at near and far distances from the Daya Bay reactors. The optical properties and as-built dimensions and positions of the vessels have been well characterized. This is a prerequisite for obtaining sub-percent systematic uncertainties in the measurement of $\theta_{13}$.

\acknowledgments

This work was done under DOE contract in the US and NSC and MOE contract in Taiwan, with support of the University of Wisconsin Foundation and National Taiwan University.  We are grateful to Reynolds Polymer Technology, Inc., of Grand Junction Colorado for their great support in engineering, fabrication, and R\&D of the outer acrylic target vessels for the Daya Bay reactor antineutrino experiment.  We are also grateful to Nakano International, Ltd, of Taiwan for their support in the construction of the inner acrylic target vessels.



\end{document}